\documentclass{emulateapj}

\slugcomment{Accepted by the Astrophysical Journal}

\shorttitle{Photophoretic Structuring}
\shortauthors{Takeuchi \& Krauss}

\begin{document}


\title{Photophoretic Structuring of Circumstellar Dust Disks}

\author{Taku Takeuchi\altaffilmark{1} and Oliver Krauss\altaffilmark{2}
}
\altaffiltext{1}{Department of Earth and Planetary Sciences, Kobe
University, Kobe 657-8501, Japan; taku@kobe-u.ac.jp}
\altaffiltext{2}{Institut f\"ur Planetologie, Westf\"alische
Wilhelms-Universit\"at M\"unster,  Wilhelm-Klemm-Str. 10, 48149
M\"unster, Germany; okrauss@uni-muenster.de}

\begin{abstract}
We study dust accumulation by photophoresis in optically thin gas
disks. 
Using formulae of the photophoretic force that are applicable for the
free molecular regime and for the slip-flow regime, we calculate
dust accumulation distances as a function of the particle size.
It is found that photophoresis pushes particles
(smaller than 10 cm) outward. For a Sun-like star, these particles are
transported to $0.1-100$ AU, depending on the particle size, and forms
an inner disk. Radiation pressure pushes out small particles ($\la 1$
mm) further and forms an extended outer disk.
Consequently, an inner hole opens inside $\sim 0.1$ AU.
The radius of the inner hole is determined by the condition that the
mean free path of the gas molecules equals the maximum size of the particles
that photophoresis effectively works on ($100 \ \micron - 10$ cm,
depending on the dust property).
The dust disk structure formed by photophoresis can be distinguished
from the structure of gas-free dust disk models, because the particle
sizes of the outer disks are larger, and the inner hole radius depends
on the gas density.
\end{abstract}



\section{Introduction}

Protoplanetary disks are composed of gas and dust.
At initial stages of their evolution, the gas of protoplanetary disks is
as massive as $10^{-3} - 10^{-1} M_{\sun}$ (Greaves 2004), and small
dust particles are
mixed with the gas, making the disks optically thick at optical
wavelengths.
As the dust particles grow, the number of small particles reduces, and at a
certain stage, the disks become optically thin (Tanaka et al. 2005;
Dullemond \& Dominik 2005).
The amount of gas also decreases, as the gas dissipates (e.g.,
Hartmann et al. 1998; Clarke et al. 2001; Takeuchi et al. 2005;
Alexander et al. 2006a,b).
At late stages of the disk evolution, the disks become gas-free
dust disks, which are observed as Vega-type disks.
During the transition from protoplanetary disks to Vega-type disks,
there may be a stage where the disks are optically thin, but their gas
component still remains.
An example is HD 141569A, which is a 5 Myr Herbig Be star (Weinberger et
al. 2000) and has an optically thin dust disk (Augereau et al. 1999;
Weinberger et al. 1999; Fisher et al. 2000; Mouillet et al. 2001; Marsh
et al. 2002; Clampin et al. 2003).
The gas component of this system has been observed (Zuckerman 1995; Dent
et al. 2005; Goto et al. 2006) and its amount is estimated to be $\la 60
\ M_{\earth}$ (Ardila et al. 2005).

Dynamics of dust particles in optically thin gas disks is of interest in
order to investigate the structure of transitional disks.
Krauss \& Wurm (2005) considered the motion of dust
particles in a gas disk that is optically thin at optical wavelengths.
A dust particle receives stellar radiation directly, and the
radiation pressure pushes the particle outward.
In addition to radiation pressure, interaction between the particle and
the surrounding gas molecules induces photophoresis, which also pushes
the particle outward (see also Wurm \& Krauss 2006; Krauss et al. 2007).
When these outward accelerations act with the gas drag on the
particle, the particle drifts outward (Takeuchi \& Artymowicz 2001).
In the outer part of the disk ($\ga 10$ AU), where the mean free path of
the gas molecules is larger than the particle size (of $\la 1$ m), the
photophoretic force is proportional to the gas density.
When a particle moves outward to the point where the gas density is
no longer high enough to induce a strong photophoretic force, the
particle's outward motion stops.
Consequently, the dust particles pile-up at a certain distance from the
star that is determined by the density profile of the gas disk.
In a gas disk with mass $\sim 0.01 \ M_{\sun}$, particles of $100 \
\micron$ to 10 cm pile-up at several tens of AU.
This spontaneous ring formation is a characteristic feature that
is caused by photophoresis in a gaseous dust disk.

Krauss et al. (2007) have demonstrated that photophoretic dust motion
may be the key process for the transition from optically thick
protoplanetary disks to optically thin circumstellar disks with
ring-like dust distributions via the stage of transitional disks with an
inner hole and a more or less sharp transition to the outer disk that is
continuously pushed outward. 
This outward dust migration can also explain the presence of material
formed close to the sun like chondrules and CAIs in asteroids of the
main asteroid belt (Wurm \& Krauss 2006) or high temperature crystalline
silicates in comets from the Kuiper belt (Petit et al. 2006; Brownlee et
al. 2006; Mousis et al. 2007). 

In this paper, we seek to investigate other characteristic structures formed by
photophoresis, especially in the inner part of the disk, but at a stage
when most of the dust has already been built into larger bodies. 
Krauss \& Wurm (2005) focus on the dust dynamics in the outer part
of the disk ($\ga 10$ \ AU) and use a formula for the photophoretic force
that is relevant for low gas densities.
In order to investigate the structure of the whole disk, we use a
photophoresis formula that can be applied to the whole
range of gas densities.
Details of the formula for the photophoretic force adopted here are
described in \S\ref{sec:drift}.
Using this formula, we find that photophoresis does not induce dust
pile-up at the innermost region of the disk ($\la 0.1$ AU), leading to
the formation of an inner hole (see \S\ref{sec:accrad}).
At the inner part of the gas disk, the gas density may be too high
that the gas disk itself is no longer optically thin to the stellar
radiation, and thus any radiation effects on the dust such as radiation
pressure and photophoresis do not work at all.
In order to investigate whether photophoresis still works at the inner
disk, in \S\ref{sec:opacity}, we calculate the optical depth of the
disk due to Rayleigh scattering of the hydrogen molecules.
In \S\ref{sec:discussion1}, the thermal relaxation time of dust particles
is estimated and is compared to the rotation periods induced by gas
turbulence and by photophoresis.
In \S\ref{sec:discussion2}, we discuss what are the characteristic features
that photophoresis forms in dust disks, and what are the observable
differences from the gas-free dust disk structures.
In Appendix A, a simple calculation deriving the
photophoretic force is described.


\section{Radial Drift of Dust Particles}
\label{sec:drift}

\subsection{Forces Acting on Dust Particles}
\label{sec:fdust}

We consider the motion of a dust particle that resides in a gas disk
around a star.
The gas disk is assumed to be optically thin in the radial
direction. This means that the dust particle directly receives the
radiation of the central star.
The stellar radiation that is absorbed or scattered by the particle
pushes it outward. This outward force, which is called radiation
pressure (Burns et al. 1979), is given by
\begin{equation}
F_{\rm rad} = \frac{\pi a^2 Q_{\rm rad} I}{c} \ ,
\label{eq:radpres}
\end{equation}
where we assume that the particle is spherical and its radius is $a$,
$Q_{\rm rad}$ is the efficiency factor of the radiation pressure and
assumed to be unity in this paper, and $c$ is the speed of light.
We assume that the particle is on the midplane of the gas disk and the
distance from the central star is $r$.
The flux of the stellar radiation at the particle position is
\begin{equation}
I = \frac{L}{4 \pi r^2} \ ,
\label{eq:radint}
\end{equation}
where $L$ is the luminosity of the central star.
Poynting-Robertson (PR) drag can be neglected comparing to gas drag.
Even in a gas disk whose mass is as small as several earth mass, gas
drag is much stronger than PR drag as discussed below in
\S\ref{sec:timesc} (see also Takeuchi \& Artymowicz 2001).

If the particle's hemisphere facing the central star becomes
hotter than the opposite hemisphere, the particle receives the
photophoretic force in addition to the radiation pressure.
Because, in general, the particle rotates with a certain speed, a
considerable temperature gradient appears only if the thermal relaxation
time of the particle is much smaller than the rotation period (see
discussion in \S\ref{sec:discussion1} below), or
if the rotation axis is aligned in the direction of the light source
(Krauss et al. 2007).
In this paper, we assume that the effect of the particle rotation
can be neglected.
In this case, the photophoretic force is directed in the radial
direction and its value can be analytically calculated. 
In Appendix A, we describe estimate of the photophoretic
force derived by a simple model of cylindrical dust particles, but 
we use the rigorous formulae that have been derived in the literature in the
following discussion.

We consider a spherical dust particle of radius $a$.
The mean temperature $T_d$ of the particle is in general different from
the surrounding gas temperature $T_g$. In this paper, we simply assume that
$T_d=T_g$. The temperature gradient inside the particle is 
assumed to be small, and then the photophoretic force is calculated by
solving linearized perturbation equation of the temperature.
In Appendix A, we discuss the cases where the above
assumptions are not valid.

If the particle radius, $a$, is much smaller than the mean free path of
the surrounding gas molecules, $l$, i.e., the Knudsen number ${\rm Kn}
\equiv l/a \gg 1$, then the photophoretic force is calculated with the
free molecular approximation (Hidy \& Brock 1967; Mackowski
1989; Beresnev et al. 1993) and is given by
\begin{equation}
F_{\rm ph, f} = \frac{ \pi \alpha J_1 P I a^2}
{3 [k_d T_d / a + 4  \varepsilon \sigma_{\rm SB} T_d^4 + \alpha P
    v_T / 2] } \ , 
\label{eq:fph_f}
\end{equation}
where $P$ is the gas pressure, $k_d$ is the thermal conductivity
inside the particle, $\varepsilon$ is the particle's emissivity of
thermal radiation and assumed to be unity, 
$\sigma_{\rm SB}$ is the Stefan-Boltzmann constant, and $v_T$ is the
mean thermal velocity of the gas molecules.
The asymmetry parameter $J_1$ represents how effectively the
incident starlight induces a temperature gradient inside the particle.
If the incident light is perfectly absorbed at the particle surface, the 
asymmetry parameter $J_1$ has the maximum value $0.5$, and it becomes
smaller as the temperature gradient becomes smaller.
The gas molecules that hit the particle are assumed to leave the
particle with the Maxwell velocity distribution of the local surface
temperature of the particle (i.e., the accommodation factor $\alpha = 1$).
In many cases, the denominator of equation (\ref{eq:fph_f}) is dominated
by the first or second terms, i.e., the temperature gradient in the particle is
determined by the internal thermal conduction or by the
radiative cooling, rather than by the thermal conduction to the gas.
In this case, the photophoretic force is approximated by
\begin{equation}
F_{\rm ph, f} \approx \frac{ \pi \alpha J_1 k_B I a^2}
{3 \sqrt{2} \sigma_{\rm mol} k_d 
( 1 + 4 \varepsilon \sigma_{\rm SB} T_d^3 a /k_d ) } {\rm Kn}^{-1} \ ,
\label{eq:fph_f_apx}
\end{equation}
where we used $P=n_g k_B T_g$ and $l=1/(\sqrt{2} n_g \sigma_{\rm mol})$,
and $n_g$ 
is the number density of the gas molecules, $k_B$ is the Boltzmann constant,
and $\sigma_{\rm mol}$ is the collisional cross section of the gas molecules.
In the free molecular regime, the photophoretic force is inversely
proportional to the gas mean free path and proportional then to the gas
density.

If the particle size, $a$, is much larger than the mean free path of the 
gas molecules, $l$, i.e., the Knudsen number ${\rm Kn}
\ll 1$, then the photophoretic force is calculated with the
slip-flow approximation (Mackowski 1989).
Mackowski (1989) calculated the photophoretic force assuming that the thermal
radiation from the particle surface can be neglected.
When the thermal radiation is taken into account, his formula is
modified to (see Appendix B for the derivation)
\begin{equation}
F_{\rm ph, s} = \frac{ 4 \sqrt{2} C_s J_1 k_B I a^2 }
{k_d \sigma_{\rm mol} d} {\rm Kn} \ ,
\label{eq:fph_s}
\end{equation}
where
\begin{equation}
d = ( 1+ 3 C_m {\rm Kn} ) \left[ \left( \frac{4 \varepsilon \sigma_{\rm
SB} T_d^3 a}{k_d} + 1 \right)
\left(1 + 2 C_t {\rm Kn} \right) + 2 \frac{k_g}{k_d} 
\right] \ ,
\end{equation}
and $C_s$, $C_m$, and $C_t$ are the coefficients for the jump conditions at
the surface and are of order of unity.
We adopt the values of $C_s=1.17$, $C_m=1.14$, and $C_t=2.18$ (Mackowski
1989).
The thermal conductivity of the gas is
\begin{equation}
k_g = \frac{15}{8} l v_T n_g k_B\ ,
\end{equation}
where we set the Prandtl number to be $2/3$.
When $d$ is approximated by unity, i.e., ${\rm Kn} \ll 1$, $4 \varepsilon
\sigma_{\rm SB} T_d^3 a / k_d \ll 1$, and $k_g / k_d \ll1$,
the photophoretic force is proportional to the gas mean free path.

To connect the above two regimes, we introduce the following formula
\begin{equation}
F_{\rm ph} = \frac{ {\rm Kn}^2 F_{\rm ph, f} + F_{\rm ph, s}}{1+{\rm Kn}^2}
 \ .
\label{eq:fph}
\end{equation}
In the limit of ${\rm Kn} \ll 1$, $F_{\rm ph}$ approaches $F_{\rm ph,s}$, 
and in the other limit (${\rm Kn} \gg 1$), it approaches $F_{\rm
ph,f}$.

If the particle rotation cannot be neglected, there are two effects we must
consider.
First, the temperature gradient in the particle becomes smaller as the
particle rotates faster.
This effectively reduces the asymmetry factor $J_1$.
In the limit of rapid rotation, the asymmetry factor $J_1$ approaches zero.
Second, the temperature gradient generally does not coincide with the
direction of the light source, i.e., the radial direction.
The photophoretic force has a component in the azimuthal direction, and
can give or take away angular momentum of the particle, depending on
the orientation of the rotation.
The particle drifts outward or inward, and the direction of the drift
reverses when the orientation of the rotation changes.
This is similar to the Yarkovsky effect acting on asteroids (Bottke et 
al. 2002), although, in the case of photophoresis, the agents of
momentum transfer are the gas molecules and not emitted photons.
The particle radial drift due to this effect is stochastic, if the
variation in the rotational orientation occurs frequently before the
particle travels a large distance in the radial direction.
Though the study of photophoretic Yarkovsky effect is important, we leave
this for future investigations, and in this paper we assume that the
photophoretic force is always directed to the radial direction.

For further discussion, it is convenient to normalize the radiative forces given above by the gravitational force
\begin{equation}
F_{G,d} = - \frac{G M m_d}{r^2} \ ,
\end{equation}
where $G$ is the gravitational constant, $M$ is the mass of the central star,
and $m_d$ is the mass of the particle.
We introduce a normalization factor
\begin{equation}
\beta=\beta_{\rm rad} + \beta_{\rm ph} \ ,
\end{equation}
where
\begin{equation}
\beta_{\rm rad}=\left| \frac{F_{\rm rad}}{F_{G,d}} \right| 
 = \frac{3 Q_{\rm rad} L}{16 \pi G M c \rho_d a} \ ,
\label{eq:beta_rad}
\end{equation}
and
\begin{equation}
\beta_{\rm ph} = \left| \frac{F_{\rm ph}}{F_{G,d}} \right| \ .
\end{equation}
In the free molecular regime (${\rm Kn} \gg 1$),
\begin{equation}
\beta_{\rm ph} = \frac{\alpha J_1 k_B L}{16 \sqrt{2} \pi G M \sigma_{\rm
    mol} k_d \rho_d l ( 1 + 4 \varepsilon \sigma_{\rm SB} T_d^3 a /k_d )
    } \ .
\label{eq:beta_ph_f}
\end{equation}
At large distances from the star, the dust temperature is so low that $4
\varepsilon \sigma_{\rm SB} T_d^3 a /k_d \ll 1$, and that
$\beta_{\rm ph}$ is independent of the particle radius $a$.

Next we consider the forces acting on the gas. The gravity acting on a unit
volume of the gas is
\begin{equation}
F_{G,g} = - \frac{G M \rho_g}{r^2} \ ,
\end{equation}
where $\rho_g$ is the mass density of the gas.
The pressure gradient force on a unit volume is 
\begin{equation}
F_{\nabla P}= - \frac{\partial P}{\partial r} \ .
\end{equation}
We introduce a normalization factor $\eta$ in a similar way to the dust
particle,
\begin{equation}
\eta = \left| \frac{F_{\nabla P}}{F_{G,g}} \right| \ .
\end{equation}

\subsection{Gas Disk Models and Dust Parameters}
\label{sec:model}

For simplicity, we assume that the gas disk has a power-law 
profile of the temperature, $T_g$, in the radial direction $r$, and is
isothermal in the vertical direction $z$, i.e., it is written as
\begin{equation}
T_g(r)=T_{g,0} r_{\rm AU}^{-q} \ ,
\end{equation}
where the subscript ``0'' denotes quantities at 1 AU, and the
non-dimensional quantity $r_{\rm AU}$ is the radius in AU.
The gas density, $\rho_g$, is also assumed to have a power-law profile in $r$
and be in a hydrostatic equilibrium in $z$, i.e.,
\begin{equation}
\rho_g (r,z) = \rho_{g,0} r_{\rm AU}^{-p_m} \exp \left( - \frac{z^2}{2 h_g^2}
\right) \ , 
\label{eq:gasdensity}
\end{equation}
where the disk scale height is $\sqrt{2} h_g$.
The isothermal sound speed of the disk is $c_s=c_0 r_{\rm AU}^{-q/2}$,
and $h_g$ is written by
\begin{equation}
h_g(r) = \frac{c_s}{\Omega_{\rm K}} = h_0 r_{\rm AU}^{(-q+3)/2} \ ,
\label{eq:gasheight}
\end{equation}
where $\Omega_{\rm K}=(G M / r^3)^{1/2}$ is the Keplerian angular
frequency at the disk midplane.
The gas surface density is
\begin{equation}
\Sigma_g (r) = \int_{-\infty}^{+\infty} \rho_g dz
= \sqrt{2\pi} \rho_0 h_0 r_{\rm AU}^{-p} \ ,
\label{eq:gassurface}
\end{equation}
where $p=p_m+(q-3)/2$.
We adopt the following fiducial parameters for the central star and the
gas temperature profile:
$M = 1 \ M_\sun$, $L = 1 \ L_\sun$, $T_{g,0}= 278 \ {\rm K}$, $h_0 = 3.33
\times 10^{-2}$ AU, and $q=\case{1}{2}$.
The mean mass of the gas molecules is $2.34 m_{\rm H}$, where $m_{\rm H}$ is
the mass of a hydrogen atom (Nakagawa et al. 1986).
The collisional cross section of the gas molecules is $\sigma_{\rm mol}=2
\times 10^{-15} \ {\rm cm}^2$ (p. 228 in Chapman \& Cowling 1970).

We consider disks in which most of small dust grains have been
removed and consequently the disks have become optically thin even in the
radial direction. The dust opacity in optical wavelength arises mainly
from particles smaller than $10 \ \micron$ (see Fig. 4 of Miyake \&
Nakagawa 1993).
In order for the disk to be optically thin, the column density of small
grains ($\la 10 \ \micron$) from the star must be smaller than $10^{-3}
\ {\rm g \ cm^{-2}}$. 
Removal of dust grains probably occurs through coagulation of the grains
into planetesimals (Weidenschilling \& Cuzzi 1993; Tanaka et
al. 2005; Dullemond \& Dominik 2005; Nomura \& Nakagawa 2006).
In such disks, the amount of the gas component has likely also reduced.
Actually most Vega-type stars do not have detectable gas components
(Liseau \& Artymowicz 1998; Greaves et al 2000; Coulson et al. 2004;
Chen \& Kamp 2004), and 
only a few objects have gas as massive as several tens earth masses.
For example, HD 141569A has {a gas mass of $\la 60 M_{\earth}$ (Zuckerman et
al. 1995; Ardila et al. 2005).
Therefore, in model A, we consider a disk in which the amount of the gas
inside 100 AU is $M_g = 2.4 \times 10^{-5} M_{\sun} = 7.9 M_{\earth}$.
(The model disk extends over 100 AU. We specify the disk mass as the
mass inside 100 AU.)
We take a surface density profile proposed by Hayashi et al. (1985) in
which the power-law index is $p=1.5$.
The gas surface density is  $\Sigma_g = 1.7 \ r_{\rm AU}^{-1.5} \ {\rm g
\ cm^{-2}}$.
In addition to model A, we also consider a more massive disk to
investigate a case in which the disk opacity due to the dust has
been significantly reduced before considerable gas dissipation occurs.
In model B, the gas amount inside 100 AU is $M_g = 1.2 \times 10^{-2}
M_{\sun}$.
Further, we assume the density profile is more gradual than model A and
$p=0.5$.
This gentle slope of the density profile assures that the gas density in
the innermost part of the disk (say $\sim 0.1$ AU) does not become so
high that the Rayleigh scattering by molecular hydrogen makes the gas
disk itself become optically thick to the starlight (see
\S\ref{sec:opacity_ray} below).
The gas surface density in model B is $\Sigma_g = 25 \ r_{\rm
AU}^{-0.5} \ {\rm g \ cm^{-2}}$.

For dust particles, we use the following fiducial parameters:
The thermal conductivity $k_d = 10^2 \ {\rm erg \ s^{-1} \ cm^{-1} \
K^{-1}}$, the particle bulk density $\rho_d = 1 \ {\rm g \ cm^{-3}}$, and
the asymmetry factor $J_1 = 0.5$, the emissivity $\varepsilon =1$, and
perfect accommodation of gas molecules $\alpha=1$.
The adopted value of the thermal conductivity is typical for porous
aggregates (Presley \& Christensen 1997).
If the dust particles are more compact, the thermal conductivity can
be much higher.

\subsection{Radial Motion of Dust Particles}
\label{sec:radialmot}

Both the radiation pressure and the photophoretic force direct outward.
If the sum of these forces is stronger than the gravity of the central
star, i.e., $\beta > 1$, then the particle is not bound to the central
star and is ejected from the system.
If $\beta < 1$, the particle is bound to the central star, but it feels
a net force weaker than the gravity of the central star, and
its orbital motion is slower than Keplerian.
Assuming the particle's orbit is circular and neglecting any contribution of
the gas drag force, the azimuthal velocity of the 
particle is 
\begin{equation}
v_{\theta,d} = (1 - \beta)^{1/2} v_{\rm K} \ ,
\label{eq:vth_d}
\end{equation}
where $v_{\rm K}=(GM/r)^{1/2}$ is the Keplerian velocity.
The azimuthal velocity of the gas is also sub-Keplerian because the
pressure gradient force usually directs outward, and is given by
\begin{equation}
v_{\theta,g} = (1 - \eta)^{1/2} v_{\rm K} \ .
\label{eq:vth_g}
\end{equation}

Now we consider the effect of the gas drag force acting on a particle
that is bound to the central star (i.e., $\beta < 1$).
If the particle orbits faster than the surrounding gas ($v_{\theta,d} >
v_{\theta,g}$), then the gas drag transfers the angular
momentum of the particle to the surrounding gas.
Consequently, the particle loses angular momentum and drifts
radially to the central star. 
From equations (\ref{eq:vth_d}) and (\ref{eq:vth_g}), it is seen that
this inward drift occurs if $\beta < \eta$.
On the other hand, if the particle orbits slower than the surrounding
gas ($v_{\theta,d} < v_{\theta,g}$, i.e., $\beta > \eta$), then the particle
gets angular momentum from the surrounding gas, and drifts radially away
from the central star.
When the particle happens to be on the orbit in which its azimuthal
velocity equals the gas azimuthal velocity, there is no angular momentum
transfer by gas drag, and the particle stays on the equilibrium orbit.
More detailed discussions including the derivation of the drift velocity were
written in Takeuchi \& Artymowicz (2001).
From their equation (26) for the drift velocity, 
neglecting the term of the Poynting-Robertson drag, $\beta_c T_s$, gives
\begin{equation}
v_{r,d}=\frac{\beta - \eta}{T_s + T_s^{-1}} v_{\rm K} \ ,
\label{eq:vrd}
\end{equation}
where the non-dimensional stopping time is 
\begin{equation}
T_s = \frac{\rho_d a v_{\rm K}}{\rho_g r v_{\rm T}}  \ ,
\label{eq:tstop}
\end{equation}
[the factor $4/3$ difference between eq. (10) of Takeuchi \&
Artymowicz (2001) and eq. (\ref{eq:tstop}) comes from the different
definition of $v_{\rm T}$, see their eq. (6)]

\begin{figure}
\epsscale{1.2}
\plotone{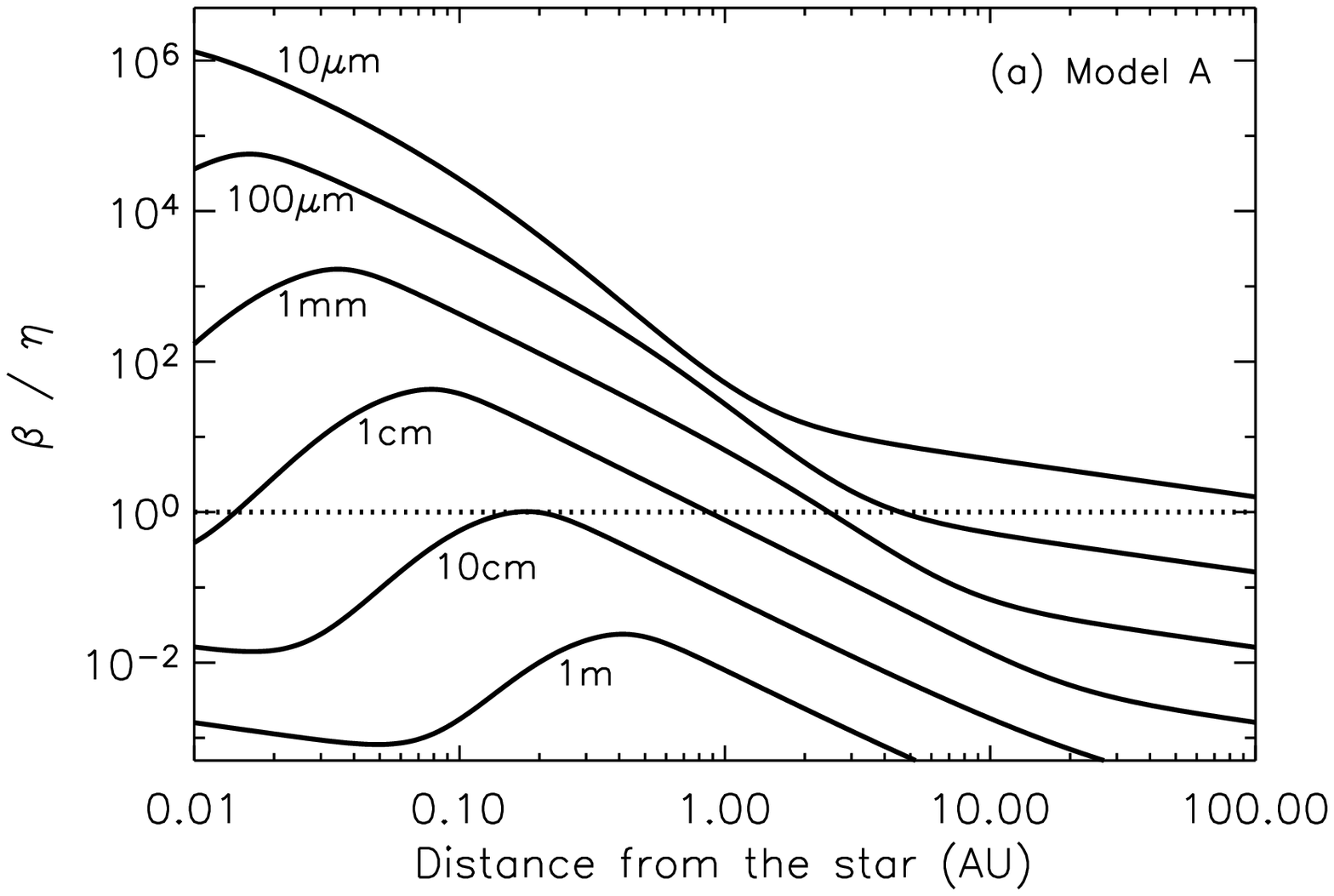}
\plotone{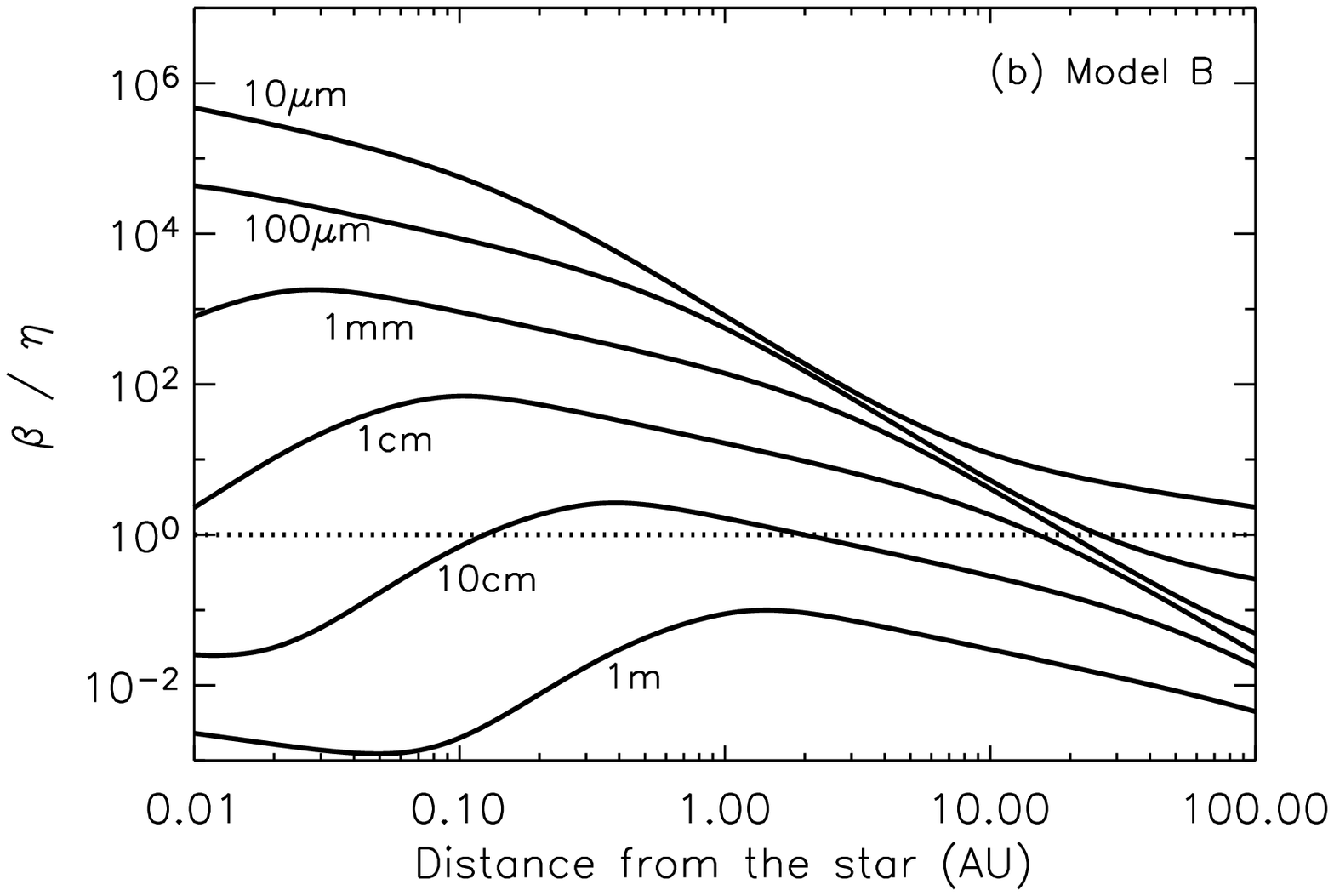}
\caption{
Ratio of the normalized outward force on the dust particle, $\beta$, with
regard to that on the gas, $\eta$.
The particle size is varied from $10 \ \micron$ to 1 m.
The dotted line represents $\beta = \eta$.
The particles drift outward if $\beta / \eta > 1$, and drift inward if
$\beta / \eta < 1$.
($a$) For the gas disk model A ($\Sigma_g = 1.7 \ r_{\rm AU}^{-1.5} \ {\rm g
\ cm^{-2}}$, $M_g = 2.4 \times 10^{-5} M_{\sun} = 7.9
M_{\earth}$ inside 100 AU). 
($b$) Model B ($\Sigma_g = 25 \ r_{\rm AU}^{-0.5} \ {\rm g \ cm^{-2}}$,
$M_g = 1.2 \times 10^{-2} M_{\sun}$).
\label{fig:beta/eta}
}
\end{figure}

We can know the direction of the particle's radial drift by seeing
if $\beta$ is larger than $\eta$ or not.
Figure \ref{fig:beta/eta}$a$ shows the ratio $\beta/\eta$ against the
distance from the central star for model A.
If a particle has $\beta$ larger than $\eta$ of the surrounding gas, it
drifts away from the star, and vice versa.
The ratio of radiation pressure to gravity, $\beta_{\rm rad}$,
which is inversely proportional to the particle radius $a$, 
is constant with the distance from the star $r$ (see
eq. [\ref{eq:beta_rad}]).
On the other hand, as long as the gas mean free path $l$ is larger than
the particle radius $a$ (free molecular regime), the photophoretic force
is inversely proportional to the Knudsen number ${\rm Kn}$ (see
eq. [\ref{eq:fph_f_apx}]), and hence decreases with $r$.
Thus, at large distances from the central star, the radiation pressure
dominates the photophoretic force, i.e., $\beta \approx \beta_{\rm
rad}$, while at small distances, the photophoretic force dominates and 
$\beta \approx \beta_{\rm ph}$.
The transition occurs at 1 AU for $10 \ \micron$ particles, 3 AU for $100
\ \micron$ particles, and 7 AU for $1 \ {\rm mm}$ particles.
The curves in Figure \ref{fig:beta/eta}$a$ bend at these distances and
the slopes become gentler at larger distances.
The slopes of the curves, $d (\beta / \eta) / dr$, are determined only by
$d \eta / dr$ at large distances, because $\beta_{\rm rad}$ is constant
with $r$, and have the same value for differently sized particles.
Inside the transition distances, where the photophoretic force
dominates, $\beta (\approx \beta_{\rm ph})$ increases with decreasing 
the Knudsen number ${\rm Kn}$, or decreasing the distance from the
central star. 
This increase in $\beta$ continues until the mean free path of the gas
molecules $l$ becomes smaller than the particle radius $a$, i.e., until
${\rm Kn}$ becomes less than unity.
At the innermost part of the disk in which ${\rm Kn} < 1$ (slip-flow
regime photophoresis), $\beta$ is proportional to ${\rm Kn}$ (see eq.
[\ref{eq:fph_s}]) and decreases with decreasing $r$.
The transition from the free molecular regime photophoresis to the
slip-flow regime photophoresis occurs at the distance where ${\rm Kn} =
1$, and at this transition distance, $\beta / \eta$ has the maximum value.
For $100 \ \micron$ particles, $\beta / \eta$ has a peak at $0.02$ AU,
and for larger particles the peak position shifts outward.
It is $0.2$ AU for $10$ cm particles.

It is seen from Figure \ref{fig:beta/eta}$a$ that particles smaller than
$10$ cm have a region where $\beta$ is greater than $\eta$.
In such a region the particles drift outward.
The outward drift of the particles stops at the locations where $\beta=\eta$.
At these locations, the particles that drift inward from the outer region
also stop.
Hence, the particles accumulate at the location where $\beta=\eta$ and $d
(\beta/\eta) /dr < 0$, leading to the formation of a dust ring as proposed
by Krauss \& Wurm (2005).
Particle accumulation does not occur at the location where $\beta=\eta$
but $d (\beta/\eta) / dr >0$, because the particles drift away from such
locations.
Note that, at the locations where particles accumulate, the Knudsen
number ${\rm Kn}$ is always larger than unity, and thus the
photophoretic force is expressed in its form for the free molecular regime.
The location of particle accumulation is at 5 AU for $100 \ \micron$
particles, and moves closer to the central star as the particle radius
increases.
Particles of 10 cm accumulate at $0.2$ AU.
If the particles are larger than 10 cm, they always drift toward the central
star and cannot accumulate anywhere.
In summary, particle accumulation occurs if $\beta/\eta>1$ somewhere in
the disk. The accumulation location is specified by the conditions
$\beta=\eta$, $d (\beta/\eta) /dr < 0$, and ${\rm Kn}>1$.

For 10 cm and 1 m particles, the curves in Figure \ref{fig:beta/eta}$a$
again bend at $0.02-0.05$ AU.
Inside these distances, the radiation pressure becomes stronger than the
photophoretic force because the Knudsen number is too low.
Thus, $\beta$ keeps constant with $r$ in this region, while $\eta$
increases, resulting in a decrease in $\beta / \eta$.

A similar result is obtained for model B (Fig. \ref{fig:beta/eta}$b$). 
In this model, particles of all sizes are driven farther outward
compared to model A.
This is because the gas density is higher in the model B disk, resulting
in an increase in $\beta_{\rm ph}$.
In  model B, it is also} apparent that $100 \ \micron$ $-$ 1 cm particles
accumulate in a narrow region of $20-30$ AU.
This pile-up of particles in a narrow ring has been pointed out by Krauss \&
Wurm (2005). 
It is the consequence of $\beta_{\rm ph}$ being independent of the
particle radius at large distances (eq. [\ref{eq:beta_ph_f}]).

Note that accumulation of outward migrating particles to the ring radius
do not increase the optical depth to the star. Formation of the dust
ring through clearing the inner dust disk does not suppress further
evolution by photophoresis. However, after clearing the inner dust, the optical
depth of the ring increases as particles from the outer disk accumulate.
When the dust ring becomes optically thick, photophoresis is weakened
and then the ring shrinks. In this paper, we consider the stage where
the dust ring is optically thin.

\begin{figure}
\epsscale{1.2}
\plotone{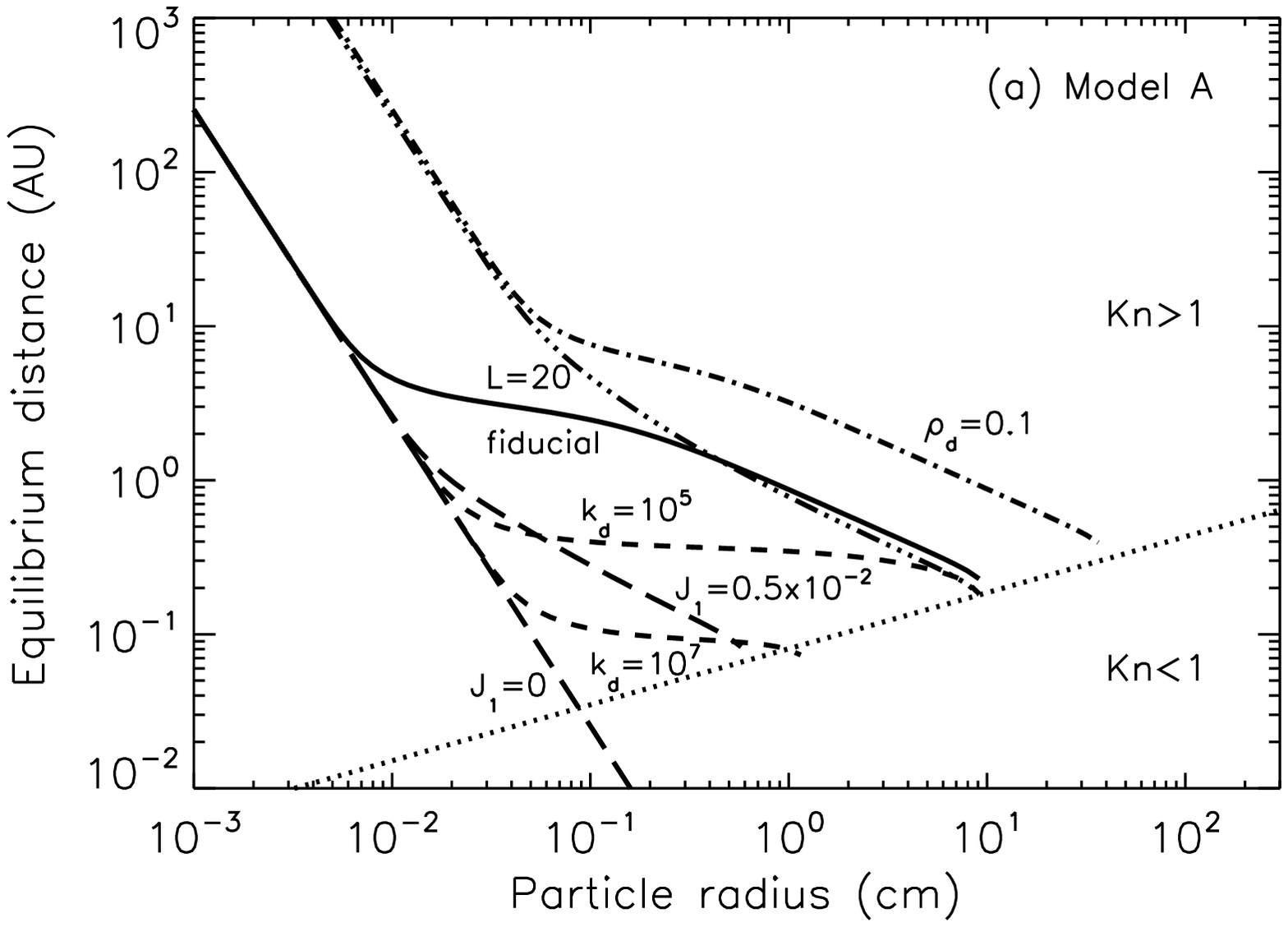}
\plotone{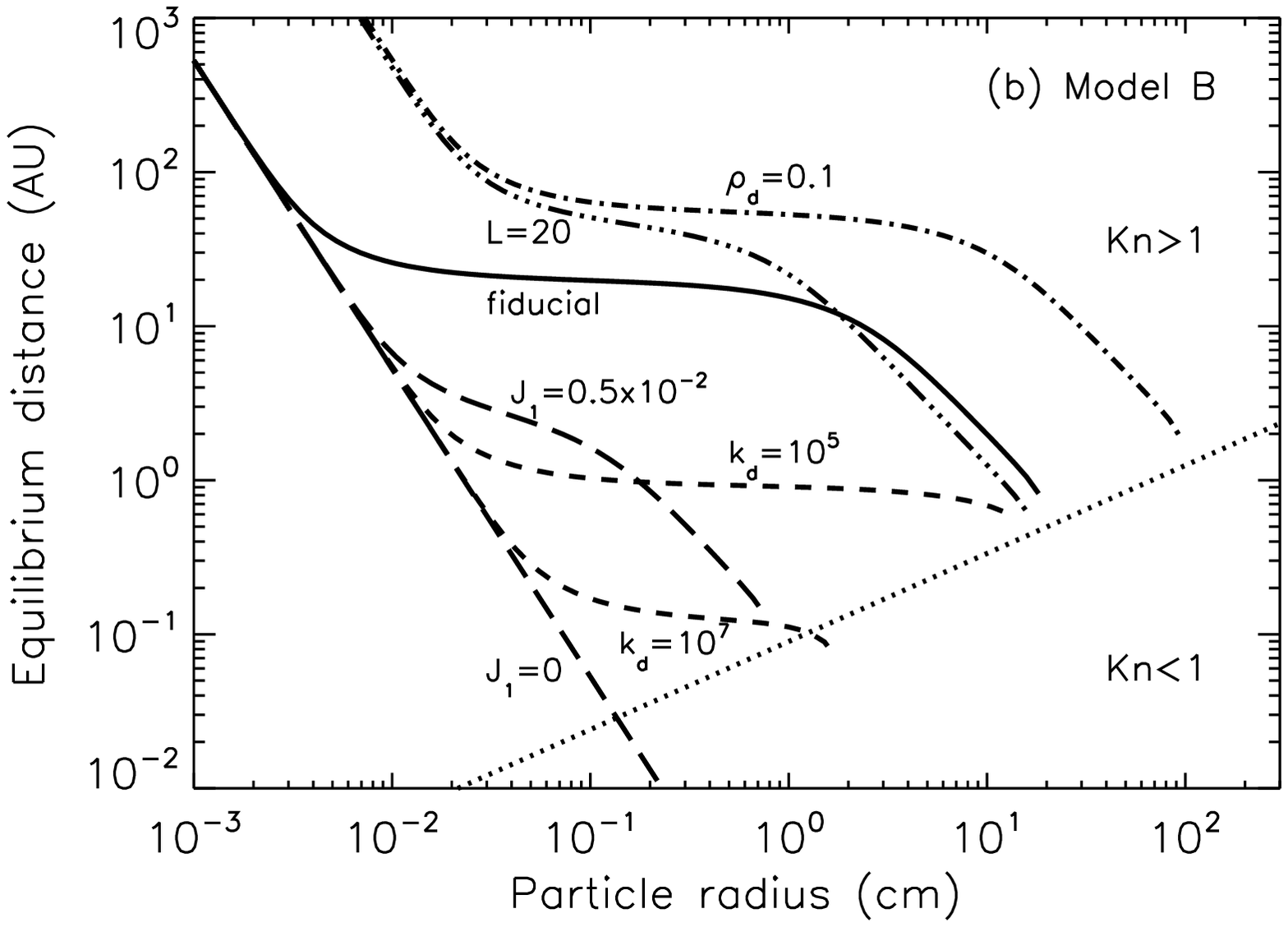}
\caption{
Equilibrium distances, where $\beta=\eta$, for the dust particles of
the various physical properties:
fiducial parameters ($k_d = 10^2 \ {\rm erg \ s^{-1} \ cm^{-1} \
K^{-1}}$, $\rho_d = 1 \ {\rm g \ cm^{-3}}$, and $J_1 = 0.5$; solid
line), 
higher thermal conductivity ($k_d = 10^5$ and $10^7 \ {\rm erg \
s^{-1} \ cm^{-1} \ K^{-1}}$; short dashed lines), smaller efficiency of
photophoresis ($J_1 = 0.5 \times 10^{-2}$ and 0; long dashed lines),
lower bulk density ($\rho_d = 0.1 \ {\rm g \ cm^{-3}}$; dot-dashed
line), and more luminous star $L=20 L_{\sun}$ (three dots-dashed line).
The dotted line shows the distances where the Knudsen number is
unity. Above (below) the dotted line, ${\rm Kn} > 1$ (${\rm Kn} < 1$).
($a$) Model A.
($b$) Model B.
\label{fig:req}
}
\end{figure}

\section{Particle Accumulation Radius}
\label{sec:accrad}

As discussed in the above section, particles accumulate at the equilibrium
locations where $\beta = \eta$ and $d(\beta / \eta) / d r < 0$.
In Figure \ref{fig:req}$a$ and \ref{fig:req}$b$ the equilibrium
distances for model A and model B, respectively, are plotted as
functions of the particle radius.
The line labeled ``fiducial'' represents the equilibrium distances
calculated with the parameters written in \S\ref{sec:model}.
The equilibrium distances for other parameters are also plotted.
We vary the value of one parameter with keeping the other parameters
being the same as the fiducial model.
Higher thermal conductivities $k_d = 10^5$ and $10^7 \ {\rm erg
\ s^{-1} \ cm^{-1} \ K^{-1}}$, a lower bulk density $\rho_d = 0.1
\ {\rm g \ cm^{-3}}$, a lower efficiency of photophoresis $J_1 = 0.5
\times 10^{-2}$, and a higher luminosity of the central star $L = 20 \
L_\sun$ are investigated.
We also calculate the case in which photophoresis does not work at
all ($J_1=0$).

\subsection{Dust Concentration and Size Segregation}

At the accumulation locations, photophoresis is under the free molecular
regime (${\rm Kn} >1$), and as seen from equation (\ref{eq:beta_ph_f}),
$\beta_{\rm ph}$ converges to a certain value in the limit of small $a$.
On the other hand, $\beta_{\rm rad}$ is inversely proportional to $a$
(neglecting the size dependence of $Q_{\rm rad}$), and thus radiation
pressure is more efficient than photophoresis for smaller particles. 
In the condition $\beta = \eta$ that determines the accumulation location,
$\beta$ is controlled by radiation pressure for small particles, and
by photophoresis for large particles.
The particle size where the transition from one behavior to the other
occurs depends on the parameters that are relevant for the strength of
the photophoretic force, i.e., $k_d$ and $J_{1}$.
This can be seen in Figure \ref{fig:req}$a$ where the curves for different
parameters depart from the straight line for $J_{1}=0$ at different
particle sizes. 
For the fiducial dust parameters in model A (the solid line in
Fig. \ref{fig:req}$a$), the accumulation locations of
particles larger than $100 \ \micron$ are controlled by
photophoresis, while those of smaller particles are controlled by
radiation pressure.

Particles that are subject to photophoresis have the tendency to
accumulate in a narrow ring region, provided that all particles have the
same physical properties, i.e., they have the same values of $\rho_d$,
$k_d$, and $J_1$.
For example, in the fiducial model (the solid line in
Fig. \ref{fig:req}$a$), particles of $200 \ \micron \ - \ 2$ mm
accumulate in a region of $2-4$ AU, and $300 \ \micron - 10$ cm
particles of the higher conductivity $k_d = 10^5 \ {\rm erg
\ s^{-1} \ cm^{-1} \ K^{-1}}$ accumulate at $0.2-0.5$ AU.
Hence, a significant influence of photophoresis is to condense the
particles of a certain size interval. 
For small particles, this size interval is limited by the increasing
influence of radiation pressure. 
For large particles, the limit is determined by the increasing
importance of thermal emission and conduction to the gas for the thermal
relaxation and the approach to ${\rm Kn}=1$.  
The location of accumulation decreases with increasing thermal
conductivity. Thus, particles with different thermal properties can be
separated in the disk by this process.  
While $k_d = 10^2 \ {\rm erg \ s^{-1} \ cm^{-1} \ K^{-1}}$
corresponds to very porous dust aggregates, values around $10^5 \ {\rm
erg \ s^{-1} \ cm^{-1} \ K^{-1}}$ are appropriate for compact rocky or
glassy grains and $10^7 \ {\rm erg 
\ s^{-1} \ cm^{-1} \ K^{-1}}$ for purely metallic particles.
As discussed by Wurm \& Krauss (2006), compact particles like chondrules
or CAIs might have been accumulated in the region of the asteroid belt
whereas porous dust aggregates were driven farther outwards by
photophoresis in the forming solar system (see Fig. \ref{fig:req}$b$ for
the disk model as massive as the solar nebula). 

In actual disks, the properties of particles are probably not
homogeneous, but the dust disks are composed of a wide variety of
particles.
Even particles of the same size therefore spread over a certain range of 
distances, and the dust ring formed by photophoresis has some
extent in the radial direction.
We still expect, however, particle concentration in the disk happens to
some degree.
Even if the parameter range of the particle properties is as wide as that
considered in Figure \ref{fig:req}$a$, we clearly see two populations of dust
particles.
All the particles at distances larger than 10 AU are
smaller than 1 mm and do not show any sign of concentration.
The dust population inside 10 AU is composed of particles larger than $100 \
\micron$, and is confined in a region between $0.1-10$ AU. We call this
region as an inner disk.
These two populations are divided by the condition which of the radiation pressure or
the photophoretic force is larger than the other.
The population of small particles is subject to radiation pressure,
and the other is subject to photophoresis.
Hence, the effect of photophoresis is to discriminate large particles
from small particles and to confine them to the inner region of the
disk.

\subsection{Inner Holes}

Another significant feature of Figure \ref{fig:req}$a$ is that in the region
below the dotted line, which represents the area where ${\rm Kn}<1$, there is
no equilibrium location where particle accumulation through 
photophoresis occurs.
As discussed in \S\ref{sec:radialmot}, particle accumulation by
photophoresis always occurs at locations where ${\rm Kn}>1$.
If photophoresis does not work ($J_1$=0), it is seen in Figure
\ref{fig:req}$a$ that particles can accumulate even in the region where
${\rm Kn} < 1$.
This feature will be a possible observational evidence for showing that 
photophoresis actually occurs in gas disks, though it 
requires observations of the innermost part of the disk ($r \la 0.1$ AU)
with a very high spatial resolution.
If photophoresis actually works, we would find that an inner hole opens
in the dust disk, and that its radius is determined by the condition
${\rm Kn}=1$.
If photophoresis does not work, on the other hand, the dust particles
would exist even in a region where ${\rm Kn}<1$.
An actual dust disk is a mixture of particles of various physical
properties. 
If the dust particles have the range of properties that are considered
in Figure \ref{fig:req}, the inner radius (that is determined by the
particles of $k_d=10^7 \ {\rm erg \ s^{-1} \ cm^{-1} \ K^{-1}}$ or
$J_1=0.5 \times 10^{-2}$) is $\sim 0.1$ AU. This inner radius may be as
small as $0.03$ AU, if there are considerable numbers of particles on that
photophoresis works very weakly (i.e., particles of very small $J_1$,
but in that case the inner disk clearing is probably suppressed as shown
by the $J_1=0$ line.)

We stress that the Knudsen number ${\rm Kn}$ is only a function of the
dust particle size $a$ and the mean free path of the gas molecules $l$.
Thus, the inner radii of dust disks do not directly reflect the dust
particles' physical properties such as the thermal conductivity $k_d$, the
bulk density $\rho_d$, and the efficiency of photophoresis $J_1$.
(The particle size at the inner edge does depend on the particle
properties.)
The fact that the inner radius of the dust disk is determined only by
the particle size considerably helps us when we observationally
test the idea of inner hole opening by photophoresis. On the other hand, it
is of great importance to have an independent measurement of the local
gas density to decide whether the condition ${\rm Kn}=1$ is fulfilled,
which might still be a challenging task for modern observational
techniques.

\subsection{Massive Disks}

Figure \ref{fig:req}$b$ shows the equilibrium distances with the same
parameter combinations as in Figure \ref{fig:req}$a$ but for the disk
model B.
In model B, where the disk mass (inside 100 AU) is $500$ times larger
than that of model A, 1 mm to 10 cm particles may extend farther
from the central star than in model A due to photophoresis (up to $\sim
100$ AU).
The inner radius of the dust disk locates at $\sim 0.1$ AU (for the
particles of $k_d=10^7 \ {\rm erg \ s^{-1} \ cm^{-1} \ K^{-1}}$ or
$J_1=0.5 \times 10^{-2}$), and is similar to that in model A.
This is just because we take such a density distribution for model B 
that the distances where ${\rm Kn}$ equals unity for the particle size
$a=1 \ {\rm mm} - 1 \ {\rm m}$ do not considerably differ from those in
model A.
This choice of the density distribution in model B comes from the
requirement that the gas disk must be optically thin and we take a
marginally optically thin disk (see \S\ref{sec:opacity_ray} below).

\subsection{Timescale of Structure Formation}
\label{sec:timesc}

\begin{figure}
\epsscale{1.2}
\plotone{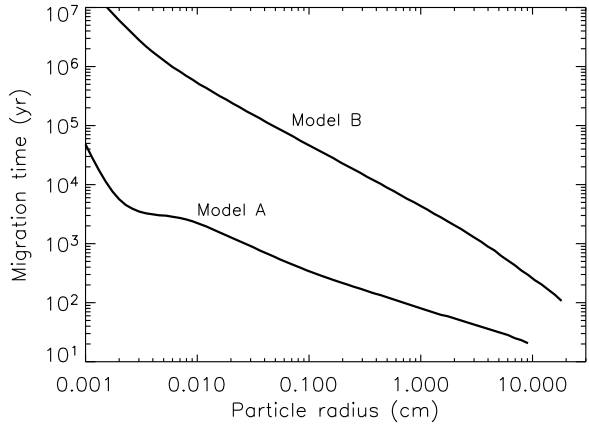}
\caption{
Timescales, $\tau_{\rm mig}$, for the dust particles to migrate to the
accumulation locations. The dust parameters are the fiducial values.
\label{fig:tmig}
}
\end{figure}

The timescale for the dust particles to migrate to the accumulation
locations is $\tau_{\rm mig} = | r_{\rm eq} / v_{r,d} |$,
where $r_{\rm eq}$ is the equilibrium distance shown in Figure
\ref{fig:req} and $v_{r,d}$ is the radial drift velocity.
In equation (\ref{eq:vrd}), $\beta \sim \eta$ near the equilibrium distance
$r_{\rm eq}$, and thus the drift velocity $v_{r,d}$ is 
estimated as $v_{r,d} \sim \eta v_{\rm K} / (T_s + T_s^{-1})$.
The migration timescale is then $\tau_{\rm mig} \sim (T_s + T_s^{-1}) /
(\eta \Omega_{\rm K})$, where the value is calculated at $r_{\rm eq}$.
Figure \ref{fig:tmig} shows the migration timescales for the fiducial
dust parameters of model A and B.
For both models, the timescales are shorter than $10^6 - 10^7$ yr, which
is the expected age of the transitional disks, except for small
particles ($a \la 100 \ \micron$) in model B.
Hence, the disk structure and the inner hole are expected to form quickly
when the disks have become transparent to the starlight.
(In disks more tenuous than model A, photophoresis effectively works
only at the dense innermost part of the disk, and thus the dust
accumulation location approaches the star. For example, in a disk
$10^{-3}$ times more tenuous than model A, i.e., in a disk of $M_g \sim 10^{-2}
M_{\earth}$, particle accumulation occurs only inside 
$0.1$ AU. The migration timescale for $100 \ \micron - 1$ cm particles
at $\la 0.1$ AU is still less than $10^4$ yr.) 

The migration timescale in Figure \ref{fig:tmig} is much shorter than the
orbital decay time due to Pointing-Robertson (PR) drag. 
For example, 100 $\micron$ particles in model A accumulate at 4 AU in
$\tau_{\rm mig} = 2 \times 10^3$ yr. The particles have
$\beta_{\rm rad}=6 \times 10^{-3}$ and the timescale due to PR drag is
$\tau_{\rm PR} = 400 (r / 1 \ {\rm AU})^2 \beta_{\rm rad}^{-1} \ {\rm
  yr} = 10^6$ yr (Burns et al. 1979), and much longer than $\tau_{\rm mig}$.
Both in model A and B, it is shown that $\tau_{\rm mig} \ll \tau_{\rm
  PR}$ for any particle size.
This means that PR drag is much weaker than gas drag and can be neglected. 

\section{Opacity of the Disk Gas}
\label{sec:opacity}

In the previous sections, we assumed that the disk is optically thin
even in the radial direction and dust particles directly receive the
light from the central star.
This assumption requires that both the dust and gas opacities of the disk
must be small enough.
We consider a stage in which most of the dust grains are packed together
into planetesimals or larger objects and the disk opacity due to
the dust has become negligible.
Such a situation has probably been realized in some Vega-type stars.
The gas disk also must be transparent, though photophoresis requires
a sufficient amount of gas.
In this subsection, we discuss whether the gas disk that is dense enough
for photophoresis to work is transparent to the starlight.

We assume that the gas disk is mostly composed of hydrogen molecules.
Thus, Rayleigh scattering by ${\rm H_2}$ is a possible and probably
dominant mechanism of light extinction in the disk.
In this subsection, we consider the gas opacity due to Rayleigh
scattering by ${\rm H_2}$ and estimate the maximum gas density at which
the gas disk is marginally opaque to the light of the central star.
Our estimate gives the lower limit of the gas opacity, because only
Rayleigh scattering by ${\rm H_2}$ is taken into account, and thus the
derived maximum gas density should be considered as an upper limit.
A transparent gas disk in which photophoresis works must be less dense
than our estimate. 
If other extinction mechanisms work more effectively than the Rayleigh
scattering by ${\rm H_2}$, this upper limit of the density becomes
smaller.
In \S\ref{sec:opacity_other} below, we briefly discuss other extinction
mechanisms.

\subsection{Rayleigh Scattering by Hydrogen Molecules}
\label{sec:opacity_ray}

The cross section of Rayleigh scattering by ${\rm H_2}$ per molecule is
\begin{equation}
\sigma_{\rm H_2} = 8.4909 \times 10^{-29} \left( \frac{\lambda}{1 \
\micron} \right)^{-4} \ {\rm cm^2} \ ,
\label{eq:sig_ray}
\end{equation}
where $\lambda$ is the wavelength of the light (Tsuji 1966).
In this subsection, we make an order of magnitude argument, and simply
estimate the opacity around the central wavelength of the central star's
spectrum, $\lambda = 0.5 \ \micron$. 
The stellar light source is assumed to be point sources, which are
located slightly above and beneath the origin, $z_* = \pm 0.424 R_*$, where
$R_*$ is the radius of the central star (Hollenbach et al. 1994) and
we assume $R_* = 1.0 \ R_{\sun}$.
The optical depth from the star, $(0,z_*)$, to a dust particle on the
midplane, $(r,0)$, is
\begin{equation}
\tau_{\rm H_2} = \int \sigma_{\rm H_2} n_g ds \ ,
\end{equation}
where $n_g$ is the number density of the gas molecules (we assume all the
gas is composed of ${\rm H_2}$ for simplicity).

\begin{figure}
\epsscale{1.2}
\plotone{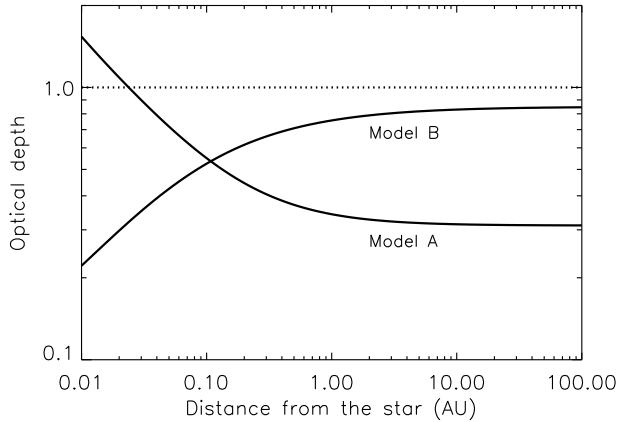}
\caption{
Optical depth $\tau_{\rm H_2}$ due to Rayleigh scattering of ${\rm
  H_2}$ from the typical point of the star $(0, 0.424 R_*)$ to the
midplane of the disk at $(r, 0)$.
The dotted line represents $\tau_{\rm H_2}=1$.
\label{fig:gasopt}
}
\end{figure}

\begin{figure}
\epsscale{1.2}
\plotone{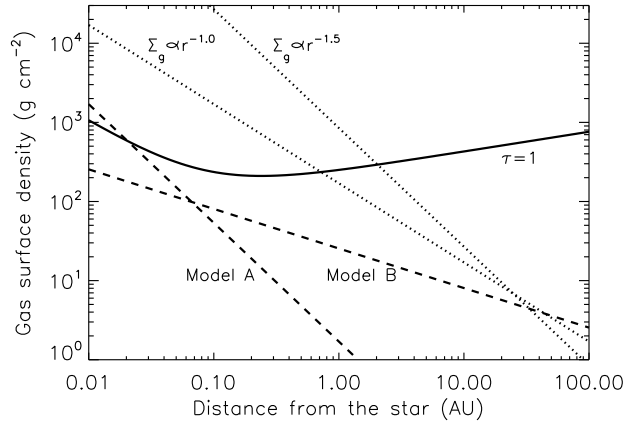}
\caption{
Comparison of the density profiles of models A and B (dashed lines) to
the maximum gas density required to be optically thin to the central
star (solid line).
The dotted lines show the density profiles that have the same amount
of mass as model B (inside 100 AU) but have different slopes ($\Sigma_g
\propto r^{-1.5}$ and $\Sigma_g \propto r^{-1.0}$).
\label{fig:gasden}
}
\end{figure}

Figure \ref{fig:gasopt} shows the optical depths from the central star as
functions of the distance $r$.
In both models A and B, the disks are optically thin (except for $r
<0.03$ AU in model A).
Note that the disk mass (inside 100 AU) of model B is $500$ times
larger than that of model A, but the optical depths from the star are not
much different.
This is because the opacities of the disks are mainly determined by
the density at $\sim 0.1$ AU and the densities at $0.1$ AU in model A and B
do not differ much (Fig. \ref{fig:gasden}).
The starlight that illuminates a particle located at a distance $r \gg
0.1$ AU gets into the disk at a distance $\sim 0.1$ AU, because the disk
scale height at $0.1$ AU is $\sqrt{2} h_g \approx 2 \times
10^{-3} \ {\rm AU} \approx 0.4 R_* \approx z_*$.
Model B has the density profile $\Sigma_g \propto r^{-0.5}$ and most
of the gas mass is distributed in the outermost part of the disk, while the
density profile of model A is more steep ($\Sigma_g \propto r^{-1.5}$)
and the disk mass is more concentrated in the innermost region.
Thus, even though the disk mass of model B is $500$ times
larger than that of model A, the densities at $0.1$ AU in model A and B
are similar.
For particles located much farther than $0.1$ AU, the optical depth
from the central star does not change with the distance $r$.

For particles inside $\sim 0.1$ AU, the location where the starlight
enters the disk moves toward the central star as the particle position
gets closer to the star.
In model A, the optical depth is determined by the gas density at where
the starlight enters the disk, which rapidly increases with decreasing
the distance $r$.
Consequently, the optical depth increases with decreasing the particle
distance.
In model B, the concentration of the gas at the innermost part of the disk
is moderate.
Thus, the decrease in the path length of the starlight (which is
approximately proportional to $r$) dominates the increase in the gas
density at where the starlight enters the disk. 
The optical depth from the star decreases with decreasing the particle
distance in model B.

We estimate the maximum gas density of a disk for it to be optically thin.
The optical depth to the star from $(r,0)$ at the disk midplane is estimated
as $\tau \approx \sigma_{\rm H_2} n_g s$, where the number density of
hydrogen molecules $n_g$ is estimated at $(r,0)$, and $s$ is the path
length of the starlight.
The path length $s$ is the distance between $(r,0)$ and
the point where the starlight enters the disk.
The location of the latter point is approximately calculated as the
point where the line connecting $(0,0)$ and $(r,\sqrt{2}h_g)$ crosses the
line connecting $(0,z_*)$ and $(r,0)$.
The condition $\tau < 1$ reduces to the condition for the gas surface
density $\Sigma_g < \sqrt{2 \pi} h_g m_{\rm H_2} / ( \sigma_{\rm H_2}
s)$, where $m_{\rm H_2}$ is the mass of a hydrogen molecule.
The solid line in Figure \ref{fig:gasden} shows the upper limit of the
gas surface density for the optical depth to the star to be less than unity.
We see that for $r>0.02$ AU both the surface densities of model A and B
(dashed lines) are less than the critical density for $\tau=1$ (solid line).
For comparison, the surface densities of the disks that have the same amount
of mass as model B (inside 100 AU) but have different slopes of the
density distribution ($\Sigma_g \propto r^{-1.5}$ and $\Sigma_g \propto
r^{-1.0}$, dotted lines).
It is seen that, if the disk is centrally concentrated, the disk is
optically thick at the inner region of the disk.

\begin{figure}
\epsscale{1.2}
\plotone{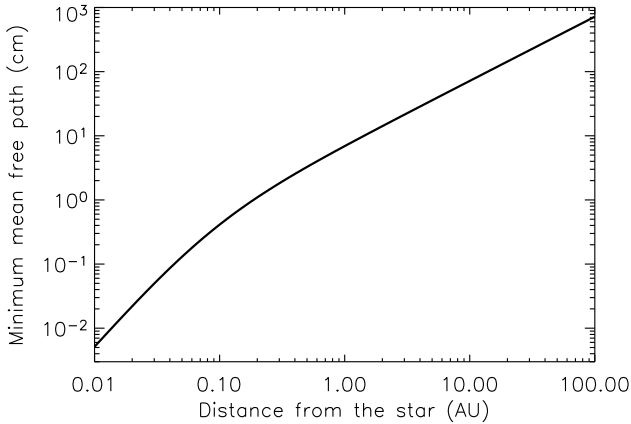}
\caption{
Mean free path of the hydrogen molecules in the critical disk where the
optical depth to the central star is unity.
\label{fig:lcri}
}
\end{figure}

The condition $\tau < 1$ reduces also to the condition for the mean free
path of the gas molecules.
The mean free path of the gas must be larger than the value shown as the
solid line in Figure \ref{fig:lcri} for the disk to be optically thin.
This minimum value is also interpreted as the minimum size of dust
particles that compose the inner edge of the dust disk,
because the location of the inner edge is determined by
the condition ${\rm Kn}=1$.
Thus, in order to observationally test whether the inner hole is produced
by photophoresis, we have to detect particle accumulation of sizes
larger than the value indicated in Figure \ref{fig:lcri}.
Suppose that, for example, a dust disk has an inner hole of radius
$0.04$ AU. If this inner hole is formed by photophoresis, the gas disk at
$0.04$ AU must be optically thin, and from Figure \ref{fig:lcri}, the
mean free path at $0.04$ AU must be larger than 1 mm. Because only
particles larger than the mean free path can accumulate through
photophoresis, the particles at the inner edge must be larger than 1 mm.
This minimum particle size becomes larger as the inner radius of the
dust disk becomes larger. 
If the inner radius is 1 AU, the particle size must be larger than 10
cm.

\subsection{Other Opacity Sources}
\label{sec:opacity_other}

1. Bound-free and free-free absorption of negative hydrogen ions ---
The cross section of bound-free and free-free absorption is proportional
to the electron pressure.
Thus, this absorption works in environments where the ionization degree is
high, and is efficient only at the innermost region of the disk,
provided that the disk gas is thermally ionized.
In both models A and B, the temperature at $0.02$ AU is $T \approx 2000$ K,
and the gas pressure is $P_{\rm H_2} \sim 10^3 \ {\rm dyn \ cm^{-2}}$
(the molecular hydrogens dominate the gas pressure).
The partial pressures of electrons and of hydrogen atoms are
$P_e / P_{\rm H_2} \sim 10^{-7}$ and $P_{\rm H} / P_{\rm H_2} \sim 3
\times 10^{-2}$, respectively (Fig. 1 of Tsuji 1966; the solar abundance
is assumed).
At the optical wavelength $\lambda = 0.5 \ \micron$, the bound-free
absorption dominates the free-free absorption, and the absorption cross
section per hydrogen molecule is 
$ \sigma_{{\rm H^-},bf+ff} \sim 10^{-23} P_e P_{\rm H} / {P_{\rm H_2}} \
{\rm cm^4 \ dyn^{-1}} 
\sim 3 \times 10^{-29} \ {\rm cm^2}$,
where we used the value of Table 2 of Tsuji (1966).
Since the cross section of the bound-free absorption sensitively depends
on $\lambda$ and $T$, we took the highest value around $\lambda
\approx 0.5 \ \micron$ and $T \approx 2000$ K in the Tsuji's Table as a
representative value (we take the value for $\lambda=0.5 \ \micron$ and
$T=1938$ K).
From equation (\ref{eq:sig_ray}), the cross section of Rayleigh
scattering for $\lambda=0.5 \ \micron$ is $\sigma_{\rm H_2}=1.36 \times
10^{-27} \ {\rm cm^2}$. 
The cross section of bound-free and free-free absorption is much smaller
than that of Rayleigh scattering even at the innermost part ($r=0.02$
AU) of disk model A and B, and thus can be ignored.
(Recall that stellar photons penetrate the disk at $\sim 0.1$ AU for
calculating the optical depth to $r \gg 0.1$ AU.)

2. Free-free absorption of negative hydrogen molecules ---
This absorption works only at the innermost part of the disks.
Thus, we estimate the cross section at $0.02$ AU of the model disks,
which is 
$ \sigma_{{\rm H_2^-},ff} \sim 10^{-27} P_e \ {\rm cm^4 \ dyn^{-1}} 
\sim 10^{-31} \ {\rm cm^2}$,
where we used equation (8) of Tsuji (1966).
This value is much smaller than that of Rayleigh scattering, and
free-free absorption of negative hydrogen molecules can be ignored.

3. Collision-induced absorption of hydrogen molecules ---
This absorption works efficiently if the gas pressure is high.
The cross section is expressed as $\sigma_{\rm CIA} = k_{\rm CIA} n_g /
n_{\rm L}^2$ where $k_{\rm CIA}$ is the opacity in the unit ${\rm
cm^{-1} \ amagat^{-2}}$ and $n_{\rm L}=2.69 \times 10^{19} \ {\rm
cm^{-3}}$ is Loschmidt's number.
We consider the innermost part of the disk.
The gas number density at $0.02$ AU of model A and B is $n_g \sim
10^{16} \ {\rm cm^{-3}}$.
At the optical wavelength $\lambda = 0.5 \ \micron$, the opacity $k_{\rm
CIA}$ due to ${\rm H_2}-{\rm H_2}$ and ${\rm H_2}-{\rm He}$ collisions
is less than $10^{-11} \ {\rm cm^{-1} \ amagat^{-2}}$ as long as $T <
4000$ K (Fig. 1 and 2 of Borysow et al. 1997). Thus, $\sigma_{\rm CIA}$
is smaller than $10^{-34} \ {\rm cm^2}$, and can be ignored in comparison
to Rayleigh scattering.

4. Thomson scattering ---
The cross section of Thomson scattering per hydrogen molecule is
$\sigma_e = 6.65 \times 10^{-25} P_e / P_{\rm H_2} \ {\rm cm^2}$.
This can be ignored in comparison to Rayleigh scattering, as long as 
$P_e / P_{\rm H_2} \ll 10^{-3}$.

5. Molecular absorption bands ---
At the innermost part of the disk, where the temperature is high enough
to vaporize refractory elements from the dust, gas molecules such as
TiO and VO contribute to the gas opacity at the optical wavelengths
(Tsuji 1971; Alexander \& Ferguson 1994; Ferguson et al. 2005).
The rotation-vibration lines of such molecules overlap and make a band
structure veiling the optical wavelengths. 
We calculated the gas opacity of model disks A and B due to the
molecular absorption bands.
In the calculation, we used the code developed by Tsuji (2002) for
calculating the opacity of cool stellar atmospheres, and assumed the
solar abundance of the gas and dust mixture.
At the optical wavelengths ($\lambda \approx 0.5 \ \micron$), absorption
by VO becomes stronger than Rayleigh scattering of ${\rm H_2}$ for $T >
1300$ K, and absorption by TiO becomes stronger for $T > 1400$ K.
Thus, in the disk inside $0.05$ AU of models A and B, where
the temperature is higher than 1300 K, the gas opacity is dominated by
the molecular absorption bands.
Our estimate of the gas opacity in \S\ref{sec:opacity_ray} is not
appropriate for the gas inside $0.05$ AU.
For dust particles at $r \gg 0.1$ AU, the rays from the central
star enter the disk at $\sim 0.1$ AU, where the gas opacity is dominated by
Rayleigh scattering, and contribution of the molecular absorption bands can
be neglected.
Most of refractory elements may be confined in large bodies such as
planetesimals and are removed from the gas phase.
In such cases, contribution of the molecular absorption bands is reduced.

\section{Discussion and Summary}

\subsection{Thermal Relaxation and Rotation Times of Dust Particles}
\label{sec:discussion1}

As mentioned in \S\ref{sec:fdust}, if particles rotate rapidly,
photophoresis does not work.
Even if particle rotation is not considerably rapid compared to the
thermal relaxation time of the particle, photophoretic Yarkovsky effect
may prevent the particle accumulation.
Such cases where photophoresis does not work are represented by the model
of $J_1=0$, where the incident starlight does not cause any temperature
gradient in the particles.
If photophoresis is suppressed by any reason, particles can reside in
the region where ${\rm Kn} < 1$ (Fig. \ref{fig:req}), and such cases can
be observationally distinguished from the disks in which photophoresis
clears the dust in the innermost region.
Hence, whether particles in gas disks rotate rapidly or not can be
observationally investigated by checking whether the dust in the region
of ${\rm Kn} < 1$ is cleared or not.

In this subsection, we estimate the thermal relaxation time of a
dust particle and then compare it to the rotation period induced by gas
turbulence or by photophoresis itself.
We use a simple model of a cylindrical dust particle described in
Appendix A.
Consider a cylindrical dust particle with radius $a$ and height $2a$
(see Figure \ref{fig:cylndust}). 
The stellar radiation flux $I$ irradiates the front surface.
We suppose that, at the beginning, the temperature inside the particle is
homogeneous.
This initial (or average) temperature $T_d$ is given by equation
(\ref{eq:td}), assuming that the incident flux $I$ balances with
thermal radiation from the front and back surfaces. (Radiation from the
side surface is ignored for simplicity.)
The temperature of the front surface increases to the equilibrium value
$T_f$ in a thermal relaxation time $\tau_{\rm th}$.
During $\tau_{\rm th}$, the incident energy at the front surface
conducts for a length $\delta a$ and forms a skin layer of temperature
gradient.
During $\tau_{\rm th}$, the temperature at the front surface is
approximated as the initial value $T_d$, and the energy flux inside the
skin layer is estimated as $Q_{\rm con} \approx I-\sigma_{\rm SB} T_d^4
= I/2$. (We set $\varepsilon=1$.)
When the skin layer depth has grown to $\delta a$ and the
temperature of the front surface converges to $T_f$, the energy flux is
$Q_{\rm con} \approx k_d (T_f - T_d) / \delta a$.
Equating the above two expressions for $Q_{\rm con}$, the thickness of the
temperature gradient layer is
\begin{equation}
\delta a = \frac{2 k_d T_d \Delta}{I} \ ,
\end{equation}
where $\Delta = (T_f - T_d)/T_d$. Note that the maximum value of $\delta
a$ is $2a$.
The thermal relaxation time is
\begin{equation}
\tau_{\rm th} = \frac{c_d \rho_d \delta a^2}{k_d} \ ,
\end{equation}
where $c_d$ is the specific heat capacity of the particle.

\begin{figure}
\epsscale{1.2}
\plotone{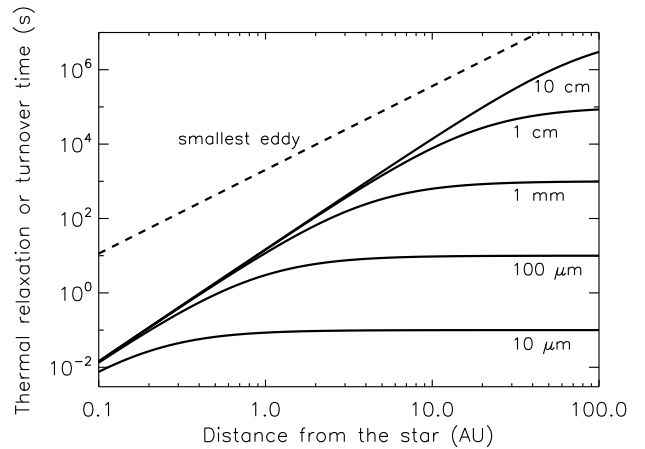}
\caption{
Thermal relaxation times of cylindrical particles of various sizes are
plotted by the solid lines against the distance from the star.
The dashed line shows the turnover time, $\tau_{\rm ed}$, of the
smallest eddies for $\alpha_{\rm tur}=10^{-2}$.
\label{fig:tauth}
}
\end{figure}

Figure \ref{fig:tauth} shows the thermal relaxation time $\tau_{\rm th}$
for $10 \ \micron - 10$ cm particles in the disk of model A.
We set the particle bulk density $\rho_d=1 \ {\rm g \ cm^{-3}}$, the
specific heat capacity $c_d=10^7 \ {\rm erg \ g^{-1} \ K^{-1}}$, and the
thermal conductivity $k_d=10^2 \ {\rm erg \ s^{-1} \ cm^{-1} \
  K^{-1}}$.
The temperature difference $\Delta$ is calculated by
equation (\ref{eq:tempdif}).
At small distances from the star, radiative cooling $4 \varepsilon
\sigma_{\rm SB} T_d^4$ dominates in determining $\Delta$ in equation
(\ref{eq:tempdif}), and $\Delta \approx 1/4$.
Thus, $\tau_{\rm th}$ is independent of the particle size $a$, and is
proportional to $T_d^{-6}$. In our model $T_d \propto r^{-1/2}$, and
thus $\tau_{\rm th} \propto r^3$.
At large distances, on the other hand, internal thermal conduction
$k_d T_d /a$ determines $\Delta$. In this case, the skin depth
is $\delta a=a$. Hence, $\tau_{\rm th}$ is proportional to $a^2$ and
independent of $r$.

If the rotation period of the particle is smaller than $\tau_{\rm th}$,
photophoresis is considerably suppressed.
Particle rotation can be excited by several mechanisms such as Brownian
motion, gas turbulent motion, collisions with other dust particles, and
the photophoretic force itself can induce rotation if it has a offset
from the mass center.
Detailed calculation of the rotation period for each mechanism is beyond the
scope of this paper.
Here, we make a rough estimate of the rotation speed induced by
turbulence and by the photophoretic force.

Suppose that the gas disk has isotropic turbulence with the Kolmogorov
energy spectrum. We assume that the largest eddies have size
$L=\alpha_{\rm tur}^{1/2} h_g$ and velocity $V=\alpha_{\rm tur}^{1/2}
c_s$, where $\alpha_{\rm tur}$ is the ``$\alpha$-viscosity'' parameter
(Cuzzi et al. 2001).
The energy of turbulent motion cascades down to smaller eddies, and
finally dissipates by the molecular viscosity.
From dimensional analysis, the energy dissipation rate per unit mass is
$\dot{e} \sim V^3 / L \sim \alpha_{\rm tur} c_s^2 \Omega_{\rm K}$.
The eddy turnover time is faster for smaller eddies and that for the smallest
eddies is
\begin{equation}
\tau_{\rm ed} \sim 
\left( \frac{\eta_{\rm vis}}{\rho_g \dot{e}} \right)^{1/2} \sim
\alpha_{\rm tur}^{-1/2} \left( \frac{l}{h_g} \right)^{1/2} \Omega_{\rm K}^{-1} \ ,
\end{equation}
where $\eta_{\rm vis}=l v_T \rho_g /2$ is the molecular viscosity (Weidenschilling 1984).
The smallest eddies can induce a particle rotation period as short as
$\tau_{\rm ed}$, if the particle is well coupled to the turbulent motion
of the smallest eddies.
However, if the particle does not strongly couple to the gas, the
rotation period is longer than $\tau_{\rm ed}$.
In Figure \ref{fig:tauth}, the turnover time, $\tau_{\rm ed}$, of the
smallest eddies is plotted by the dashed line for $\alpha_{\rm tur} =
10^{-2}$. 
The particle rotation period induced by gas turbulence is expected to be
larger than the dashed line, and therefore is much longer than the
thermal relaxation time.

\begin{figure}
\epsscale{1.2}
\plotone{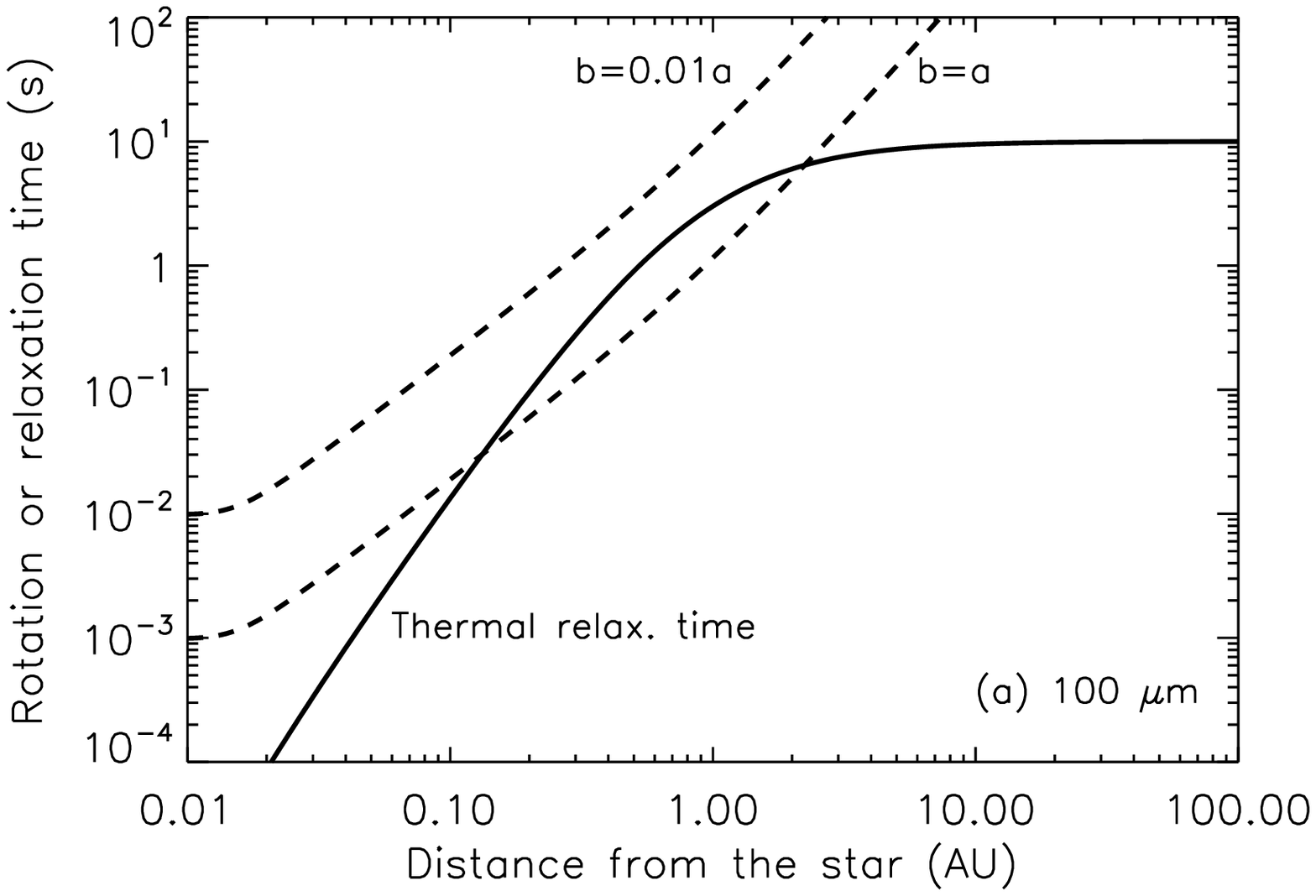}
\plotone{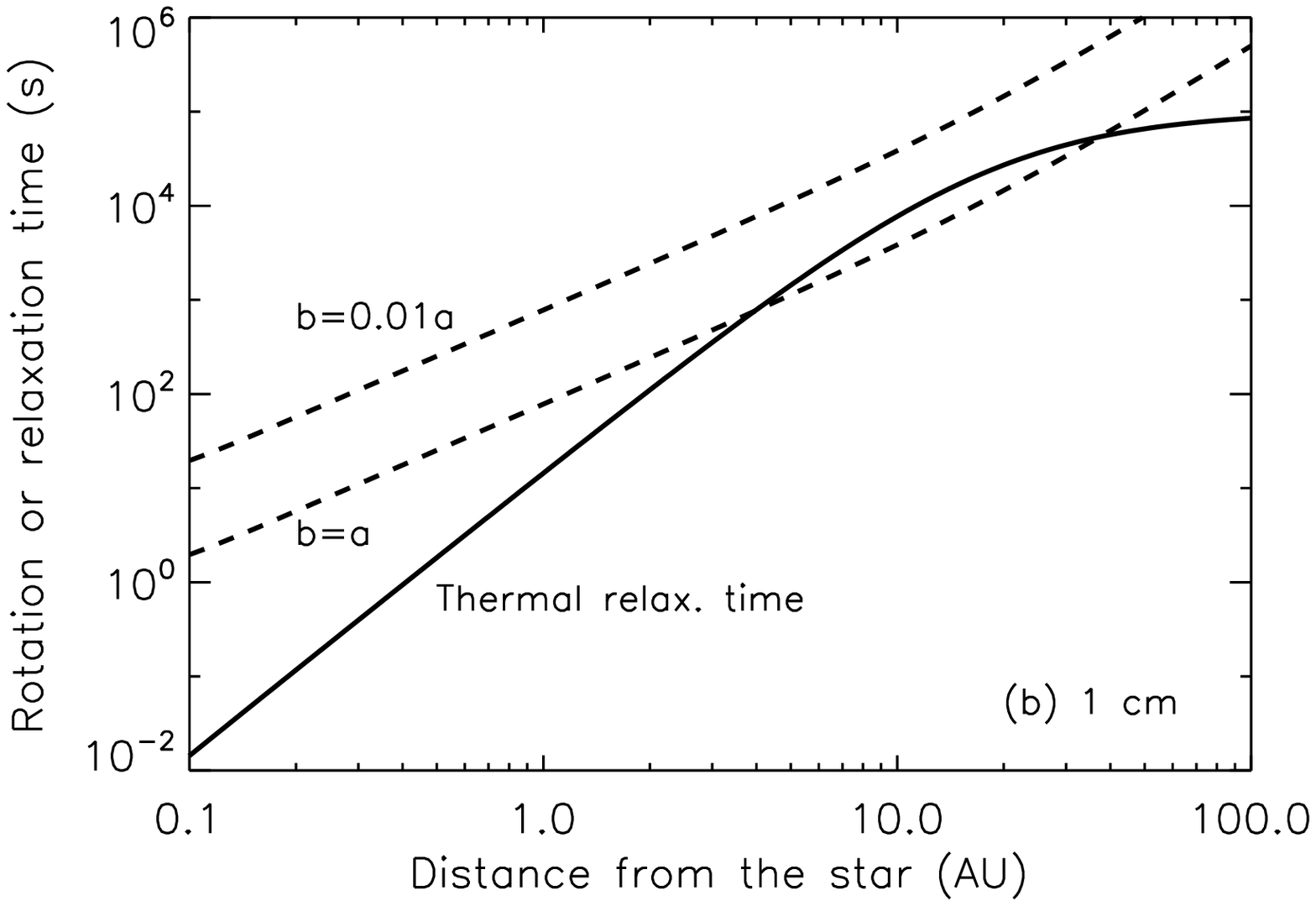}
\caption{
Times taken for a particle to rotate 180 degree by photophoresis are
plotted by the dashed lines. The photophoretic force is exerted on a
point at a distance $b=a$ or $b=0.01a$ from the cylinder axis. The solid
line shows the thermal relaxation time.
($a$) For $100 \ \micron$ particles.
($b$) For 1 cm particles.
\label{fig:rot}
}
\end{figure}

We next consider rotation induced by the torque exerted by
photophoresis.
Consider again a cylindrical dust particle and suppose that the
photophoretic force is exerted on a point at a distance $b$ from the
cylinder axis. The torque is $K=b F_{\rm ph}$. 
The value of $b$ is unknown, and we treat $b$ as a free parameter.
The principal moment of inertia of the cylinder (for the axis
perpendicular to the cylinder axis) is $I_{xx}=7 \pi a^5 \rho_d / 6$.
The time needed for a initially stationary particle to rotate 180
degree is
\begin{equation}
\tau_{\rm rot} = \left( \frac{2 \pi I_{xx}}{K} \right)^{1/2} = 
\left( \frac{7 \pi^2 \rho_d a^5}{3 b F_{\rm ph}} \right)^{1/2} \ .
\end{equation}
In Figure \ref{fig:rot}, the rotation time $\tau_{\rm rot}$ is plotted
for assumed off-centers $b=a$ and $b=0.01a$.
If the photophoretic force has a large off-center ($b=a$), the
rotation time can be shorter than the thermal relaxation time at $0.1 -
2$ AU for $100 \ \micron$ particles and at $4-40$ AU for 1 cm
particles.
In such regions, photophoresis is probably significantly suppressed.
In order for photophoresis to work effectively in the whole disk,
the off-center of the photophoretic force has to be as small as $b=0.01a$.

\subsection{Characteristic Structure Due to Photophoresis}
\label{sec:discussion2}

The dust disk structure that photophoresis makes has three zones: the
outer disk, the inner disk, and the inner hole.
The outer disk is composed of small ($a \la 1$ mm or $\beta \ga 0.01$)
particles, and their dynamics is controlled mainly by radiation pressure.
The inner disk is composed of large ($a \ga 100 \
\micron$ or $\beta \la 0.01$) particles accumulating there due to
photophoresis.
The boundary between the outer and inner regions is at $10-100$
AU, depending on the gas density profile.
The inner hole opens inside $\sim 0.1$ AU, where the Knudsen number
${\rm Kn}$ is smaller than unity for particles that photophoresis
effectively works on ($a=100 \ \micron - 10$ cm).

The structure formed by photophoresis should be compared to the
structure of gas-free disks.
Modeling of gas-free dust disks also shows zonal structure that consists
of the outer extended disk, the inner disk, and the inner hole
(Th\'ebault \& Augereau 2005; Wyatt 2006; Strubbe \& Chiang 2006; Krivov
et al. 2006).
In their models, a planetesimal belt is assumed and dust particles are
continuously produced by planetesimal collisions.
The inner disk is composed of large particles whose orbits are hardly
affected by radiation pressure and are nearly circular.
Thus, the location and width of the inner disk are basically similar to
those of the planetesimal belt.
Outside the planetesimal belt, the outer disk extends and it is 
composed of small particles whose $\beta$-value is large.
Their orbits are strongly influenced by radiation pressure and are
excited to highly eccentric orbits.
In the model of Th\'ebault \& Augereau (2005), the particles of the outer
disk have $\beta > 0.05$, which is relatively larger than $\beta$ of our
models.
In our models, particles' $\beta$ of the outer disk can be as small as
$0.01$.
Therefore, infrared to millimeter radio observations of outer disks and
determination of the particle size will provide key information for
determining which model is plausible.

Our models of photophoresis also predict that an inner hole opens in the dust
disk, but the gas still fills the hole.
This is a significant feature, but it is difficult to test this
with the present observational techniques.
We have to observe the gas, but the amount of the gas of optically thin
disks is probably quite small, and most of Vega-type stars do not have
gas in an amount exceeding the current detection limit.
Further, the radius of the inner hole is of the order of $0.1$ AU, and the
temperature there is close to the sublimation temperature of the dust.
(The temperature exceeds 1500 K inside $0.03$ AU of our model disks.)
Therefore, we need a careful observation that can distinguish between
the inner holes made by photophoresis and by dust sublimation.

\subsection{Summary}

Dust accumulation by photophoresis is studied.
We use formulae of the photophoretic force that are applicable for the free
molecular regime and for the slip-flow regime.
The main results are as follows.

1. Particle accumulation occurs at a point where the outward
   acceleration on the gas by the pressure gradient equals to the
   outward acceleration on the particle by radiation pressure and
   photophoresis.

2. Photophoresis makes an inner disk composed of relatively large
   particles ($a= 100 \ \micron - 10$ cm). The inner disk extends
   between $0.1$ AU and $10-100$ AU, surrounded by the outer disk
   composed of small particles ($a \la 1$ mm).

3. An inner hole opens inside $\sim 0.1$ AU. The inner hole radius is
   determined by the condition ${\rm Kn}=1$ for the maximum size
   particles that photophoresis effectively works on ($a=100 \ \micron -
   10$ cm). 

Photophoresis works effectively only when the disk is optically
thin. Most of small ($\la 10 \ \micron$) dust grains must be
removed from the disk such that their column density to the star becomes
smaller than $10^{-3} \ {\rm g \ cm^{-2}}$. For example, at 1 AU, the
dust density  must be smaller than $10^{-16} \ {\rm g \ cm^{-3}}$,
i.e., $10^{-5}$ times smaller than the value of the minimum mass solar
nebula model (Hayashi et al. 1985).
The gas disk also must be optically thin. Figure \ref{fig:gasden} shows
that the gas surface density must be smaller than $10^2 - 10^3 \ {\rm g
  \ cm^{-2}}$. At $0.1$ AU, where the typical ray from the star enters
the disk (for the dust particles at $r \gg 0.1$ AU), this value is $\sim
10^{-2}$ times smaller than the value of the minimum mass solar nebula
model (Hayashi et al. 1985).
Even in such a tenuous gas disk, the photophoretic force is strong enough to
change the dust disk structure inside a few AU, as shown by model A (Figure
\ref{fig:req}$a$).
If the gas density is more tenuous than model A, the region where
photophoresis has a substantial effect shrinks toward the central star.
In a gas disk with $0.1$ times smaller gas density than model A,
photophoresis on millimeter sized particles works effectively 
only inside 1 AU, and in a disk with $10^{-2}$ times smaller gas
density, the effective region shrinks to $0.3$ AU.

\acknowledgements
We thank Ingrid Mann, Tadashi Mukai, Yoichi Itoh, and Yoshitsugu Nakagawa for
useful discussions.
We are grateful to Takashi Tsuji for extensive and helpful discussions
and for kindly providing his numerical code to calculate the gas
opacity.
We are grateful to anonymous referees for their constructive comments.
This work was supported by the 21st Century COE Program, ``The Origin
and Evolution of Planetary Systems,'' of the Ministry of Education,
Culture, Sports, Science, and Technology (MEXT) of Japan, and also by the
Grant-in-Aid for Scientific Research, No. 17740107 and No. 17039009, of the
MEXT.
OK is grateful to Gerhard Wurm and acknowledges the support of the
Deutsche Forschungsgemeinschaft. 

\appendix

\section{A Simple Estimate of the Photophoretic Force \label{sec:ap-est}}

In this Appendix, we estimate the magnitude of the
photophoretic force using a simple calculation.

\begin{figure}
\epsscale{0.5}
\plotone{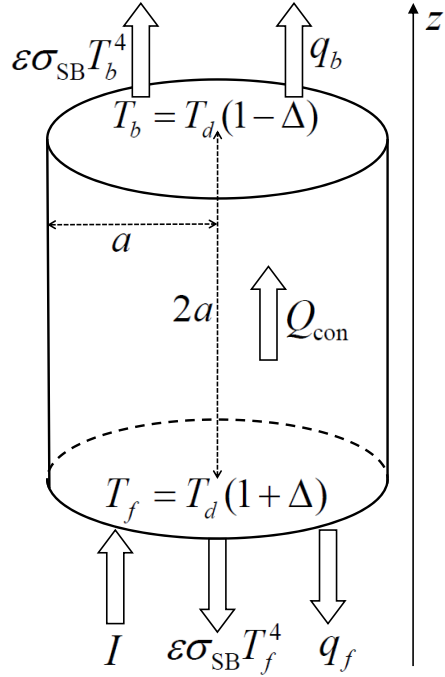}
\caption{
A cylindrical dust particle of radius $a$ and height $2a$. Incident
stellar flux $I$ is parallel to the cylinder axis ($z$-axis) and
irradiates the front surface. Radiation fluxes from the front and back
surfaces are $\varepsilon \sigma_{\rm SB} T_f^4$ and $\varepsilon
\sigma_{\rm SB} T_b^4$, respectively. The fluxes of heat conduction to
the gas are $q_f$ and $q_b$, and the heat flux inside the particle is
$Q_{\rm con}$.
\label{fig:cylndust}
}
\end{figure}

We consider a cylindrical dust particle with radius $a$ and height $2a$
(see Figure \ref{fig:cylndust}).
The particle is surrounded by the gas of temperature $T_g$.
In general, $T_g$ differs from the average temperature of the particle
$T_d$.
In deriving the photophoretic force below, we assume $T_g=T_d$, and
consider $T_g \neq T_d$ cases later.
The incident radiation flux, $I$, is parallel to the cylinder axis
($z$-axis) and irradiates the front surface.
The temperature at the front surface increases to $T_f$ and it radiates
energy flux $\varepsilon \sigma_{\rm SB} T_f^4$, where $\varepsilon$ is
the emissivity.
Some of the incident energy is conducted to the gas with a rate $q_f$
per unit area, and the residual energy flows to the back surface with a rate 
$Q_{\rm con}$ per unit area.
For simplicity, we ignore the energy transfer through the side surface.
The temperature of the back surface is $T_b$. The back surface radiates
energy flux $\varepsilon \sigma_{\rm SB} T_b^4$, and conducts heat to
the gas with a rate $q_b$ .
In an equilibrium state, the temperature inside the particle varies
linearly along the cylinder axis, and thus its values at the front and
back surfaces are written as  
\begin{eqnarray}
T_f & = & T_d (1+\Delta) \ , \\ 
T_b & = & T_d (1-\Delta) \ ,
\end{eqnarray}
respectively.
The energy balance reads
\begin{equation}
I - \varepsilon \sigma_{\rm SB} T_f^4 - q_f = Q_{\rm con} 
= \varepsilon \sigma_{\rm SB} T_b^4 + q_b \ .
\label{eq:enebla}
\end{equation}

We calculate each term of equation (\ref{eq:enebla}).
First we consider free molecular photophoresis (${\rm Kn} \gg 1$).
The radiative fluxes from the front and back surfaces are, assuming $\Delta
\ll 1$,
\begin{eqnarray}
\varepsilon \sigma_{\rm SB} T_f^4 & = & \varepsilon \sigma_{\rm SB} T_d^4 (1+4\Delta) \ , \\ 
\varepsilon \sigma_{\rm SB} T_b^4 & = & \varepsilon \sigma_{\rm SB} T_d^4 (1-4\Delta) \ ,
\end{eqnarray}
respectively.
The thermal conduction to the gas at the front surface is divided as
\begin{equation}
q_f=q_f^- - q_f^+ \ ,
\end{equation}
where $q_f^-$ is the energy loss from the particle that is taken by
the gas molecules ejected from the front surface, and $q_b^+$ is the
energy gain that is given by the adsorbed molecules.
In the free molecular approximation (${\rm Kn} \gg 1$), the velocity
distribution of the adsorbed molecules is written as $\alpha
f^+(\mbox{\boldmath$v$})$, where $0 \le \alpha \le 1$ is the
accommodation coefficient, $f^+(\mbox{\boldmath$v$})$ is the
Maxwellian velocity distribution of the temperature $T_g$ and number
density $n_g$,
\begin{equation}
f^+(\mbox{\boldmath$v$}) =n_g \left(\frac{m_g}{2 \pi k_{\rm B} T_g}
\right)^{3/2}
\exp \left( - \frac{m_g | \mbox{\boldmath$v$} |^2}{2 k_{\rm B} T_g}
\right) \ ,
\end{equation}
and $m_g$ is the mass of a molecule.
We define $f^+(\mbox{\boldmath$v$})$ to have non-zero values even for
particles going away from the surface ($v_z >0$).
The energy gain is
\begin{equation}
q_f^+ = \alpha \int_{v_z<0} \frac{1}{2} m_g v_z |\mbox{\boldmath$v$}|^2
f^+ d \mbox{\boldmath$v$} = \frac{1}{2} \alpha P v_T \ ,
\end{equation}
where $v_z$ is the velocity component normal to the front surface, $P =
  \pi \rho_g v_T^2 /8$ is the gas pressure, and $v_T = \sqrt{8 k_B T_g 
  / \pi m_g}$ is the mean thermal speed of gas molecules.
The molecules ejected from the front surface are assumed to have a
Maxwellian velocity distribution, $\alpha f^-(\mbox{\boldmath$v$})$, of the
temperature $T_f$. 
The energy loss is
\begin{equation}
q_f^- = \alpha \int_{v_z>0} \frac{1}{2} m_g v_z |\mbox{\boldmath$v$}|^2
f^- d \mbox{\boldmath$v$} = \frac{\pi}{16} \alpha
\rho_f^- (v_f^-)^3 \ ,
\end{equation}
where $v_f^- = \sqrt{8 k_B T_f / \pi m_g}$ and the ``density'' of
ejected molecules (for $\alpha=1$) is
\begin{equation}
\rho_f^- = \int_{{\rm all} \ \mbox{\boldmath$v$}}
 m_g f^- d \mbox{\boldmath$v$} \ . 
\end{equation}
From mass conservation between the adsorbed and ejected molecules
($\alpha \rho_f^- v_f^- = \alpha \rho_g v_T$), 
\begin{equation}
\rho_f^- = \left( \frac{T_g}{T_f} \right)^{1/2} \rho_g \ .
\end{equation}
Then, we obtain
\begin{equation}
q_f^- = \frac{\alpha P v_T T_f}{2 T_g}  \ .
\end{equation}
If the gas temperature and the average dust temperature are same ($T_g=T_d$),
the thermal conduction to the gas is 
\begin{equation}
q_f = \frac{1}{2} \alpha P v_T \Delta \ .
\end{equation}
Similarly, the thermal conduction to the gas from the back surface is
\begin{equation}
q_b = - \frac{1}{2} \alpha P v_T \Delta \ .
\end{equation}
The thermal conduction inside the particle is
\begin{equation}
Q_{\rm con} = k_d \frac{T_d \Delta}{a} \ .
\end{equation}
Collecting all the above expressions of the terms in equation
(\ref{eq:enebla}), the particle temperature is solved as
\begin{equation}
T_d = \left( \frac{I}{2 \varepsilon \sigma_{\rm SB}} \right)^{1/4} \ ,
\label{eq:td}
\end{equation}
and
\begin{equation}
\Delta = \frac{I}
{2 [k_d T_d / a + 4  \varepsilon \sigma_{\rm SB} T_d^4 + \alpha P v_T /
    2] } \ .
\label{eq:tempdif}
\end{equation}
The force exerted on the front surface by the ejected molecules is
\begin{equation}
F_f = \pi a^2 \alpha \int_{v_z>0} m_g v_z^2 f^- d \mbox{\boldmath$v$} 
= \frac{\pi a^2 \alpha P}{2} \left( \frac{T_f}{T_g} \right)^{1/2} \ ,
\end{equation}
and on the back surface, 
\begin{equation}
F_b = \frac{\pi a^2 \alpha P}{2} \left( \frac{T_b}{T_g} \right)^{1/2} \ .
\end{equation}
Assuming $T_g=T_d$ and $\Delta \ll 1$, the photophoretic force is
\begin{equation}
F_{\rm ph,f} = F_f - F_b = \frac{\pi a^2 \alpha I P}
{4 [k_d T_d / a + 4  \varepsilon \sigma_{\rm SB} T_d^4 + \alpha P v_T /
    2] } \ ,
\label{eq:fph_f_c}
\end{equation}
which can be compared to equation (\ref{eq:fph_f}) for the spherical
particle case.

We consider how the photophoretic force changes from equation
(\ref{eq:fph_f_c}) when the dust temperature $T_d$ differs from the gas
temperature $T_g$.
For simplicity, we consider the cases where the thermal conduction to the
gas can be neglected. 
With $q_f = q_b = 0$ in equation (\ref{eq:enebla}), the photophoretic
force becomes
\begin{equation}
F_{\rm ph,f} = \frac{\pi a^2 \alpha I P}
{4 [k_d T_d / a + 4  \varepsilon \sigma_{\rm SB} T_d^4 ] } \left(
\frac{T_d}{T_g} \right)^{1/2} \ ,
\label{eq:fphf_ap}
\end{equation}
which differs from the $T_d=T_g$ case by a factor $(T_d / T_g)^{1/2}$.
We consider further the cases where the temperature variation inside the
dust particle is large ($\Delta \sim 1$). Such a situation arises if
$\varepsilon \sigma_{\rm SB} T_d^4 \gg k_d T_d /a$, $\alpha P v_T$.
In this case, $T_f = (I / \varepsilon \sigma_{\rm SB})^{1/4}$ and $T_b$=0, and thus 
the photophoretic force is $F_{\rm ph,f}=F_f=\pi a^2 \alpha P (T_f / T_g)^{1/2}
/2$.
The photophoretic force calculated by equation (\ref{eq:fphf_ap}), which
assumes $\Delta \ll 1$, reduces to $F_{\rm ph,f}=F_f / (4 \cdot
2^{1/8})$, when the term $k_d T_d /a$ is neglected.
Thus, in the extreme case where $\Delta \sim 1$, equation
(\ref{eq:fphf_ap}) underestimates the photophoretic force by a factor $4
\cdot 2^{1/8}$.

In the slip-flow regime (${\rm Kn} \ll 1$), thermal conduction to the
gas is estimated as $q_f \sim q_b \sim k_g T_d \Delta / a$, where $k_g$
is the thermal conductivity of the gas.
The temperature difference is 
\begin{equation}
\Delta \sim \frac{I}
{[(k_d+k_g) T_d / a + 4  \varepsilon \sigma_{\rm SB} T_d^4 ] } \ .
\end{equation}
Because of the temperature variation along the dust surface, a slip
flow of the gas arises. The velocity of the thermal slip is
\begin{equation}
v_s \sim C_s v_T \Delta {\rm Kn} \ ,
\end{equation}
where $C_s \sim 1$ is the thermal slip coefficient (Lifshitz \&
Pitaevskii 1981).
The particle, which moves in the gas with a velocity $v_s$, experiences
Stokes gas drag,
\begin{equation}
F_{\rm Stokes} \sim a \eta_{\rm vis} v_s \,
\end{equation}
where $\eta_{\rm vis} = l v_T \rho_g / 2 = v_T m_g /(2 \sqrt{2}
\sigma_{\rm mol})$ is the molecular viscosity. ($F_{\rm Stokes} = 6 \pi a
\eta_{\rm vis} v_s$ if the particle is spherical.)
In the equilibrium state, the photophoretic force balances with the
Stokes drag force, and is estimated as
\begin{equation}
F_{\rm ph,s} \sim \frac{C_s a^2 k_B I}{\sigma_{\rm mol} k_d}
\left( 1 + \frac{4 a \varepsilon \sigma_{\rm SB} T_d^3}{k_d} +
  \frac{k_g}{k_d} \right)^{-1} {\rm Kn} \ ,
\end{equation}
which can be compared to equation (\ref{eq:fph_s}).
In the above derivation, we neglect the temperature jump at the surface and
the velocity slip due to imperfect sticking of the gas to the surface
($C_t=C_m=0$).

\section{Slip-Flow Regime Photophoresis}

In this Appendix, we derive the expression for the photophoretic force
in the slip-flow regime, equation (\ref{eq:fph_s}), taking into account
the radiative cooling of the particle.
We follow the derivation described in Mackowski (1989). (The assumption
is as follows. The mean temperatures of the gas and the dust, $T_g$ and
$T_d$, are the same. The temperature variation from the mean value
is small. The Reynolds number is much less than unity and thus gas drag obeys
the Stokes law. The gas temperature is determined only by the heat conduction
from the dust and inside the gas.)
Here, we do not rewrite Mackowski's derivation again, but we describe what is
modified when radiative cooling is taken into account.
In this Appendix, the numbering of equations with ``M'' means that they
refer to equations of Mackowski (1989). For the meaning of the symbols
see Mackowski (1989).
However, for some variables, we have used different symbols in the main
text, and we keep the same notation in the Appendix. When we use a
different symbol from Mackowski (1989), the corresponding symbol in
Mackowski (1989) is given in brackets.

The basic equations (M11)-(M14) and the boundary conditions (M15)-(M20)
are not changed except that equation (M16) is modified as
\begin{equation}
- k_g \frac{\partial \hat{T_g}}{\partial r}  + \varepsilon \sigma_{\rm SB}
\hat{T_d}^4 = - k_d \frac{\partial \hat{T_d}}{\partial r} \ ,
\end{equation}
where $\hat{T_d} (T_s)$ is the temperature inside the dust particle,
$\hat{T_g} (T_g)$ is the gas temperature, and $k_d (k_s$) is the thermal
conductivity of the dust. 
Then, equation (M22) becomes
\begin{equation}
D_n = \frac{n G_n(1)-G_n'(1)}
{(4 \varepsilon \sigma_{\rm SB} T_d^3 a / k_d + n) 
[1+(n+1) C_t l / a] + (n+1)k_g / k_d } \ ,
\end{equation}
where $T_d (T_0)$ is the mean temperature of the dust, $C_t (c_t)$
is the coefficient of the jump condition, and we correct the typo in the
sign of the original equation.
Equation (M28) is not modified when it is expressed using $D_1$ as
\begin{equation}
c_1 = \frac{3 (1 + 2 C_m l / a)}{4 (1 + 3 C_m l / a)}
- \frac{C_s \eta_{\rm vis} D_1}{2 V_0 \rho_g a (1 + 3 C_m l / a)} \ ,
\end{equation}
where $\eta_{\rm vis} (\eta)$ is the molecular viscosity, and $C_m
(c_m)$ and $C_s (c_s)$ are the coefficients of the jump conditions.
Finally equation (M29) becomes
\begin{equation}
F_p = - \frac{4 \pi C_s \eta_{\rm vis}^2 a I J_1}{k_d T_d \rho_g}
\frac{1}{(1 + 3 C_m l / a) [(4 \varepsilon \sigma_{\rm SB} T_d^3 a / k_d
+ 1) (1 + 2 C_t l / a) + 2 k_g / k_d ] } \ ,
\end{equation}
where $I (I_{\lambda})$ is the incident flux of the starlight.
Substitution of the expression for $\eta_{\rm vis}^2= \sqrt{2} l \rho_g k_B T_d /(\pi
\sigma_{\rm mol})$ and taking the absolute value give equation
(\ref{eq:fph_s}).




\end{document}